\documentclass[twocolumn]{aastex61}

\newcommand{\vect}[1]{\textbf{\textit{#1}}}

\received{2017 July 20}
\revised{2017 August 14}
\accepted{2017 August 14}
\submitjournal{ApJ}

\begin{document}

\title{Magnetic Flux Rope Identification and Characterization from Observationally-Driven Solar Coronal Models}

\correspondingauthor{Chris Lowder}
\email{chris.lowder@durham.ac.uk}

\author[0000-0001-8318-8229]{Chris Lowder}
\affil{Department of Mathematical Sciences \\
Durham University \\
Durham DH1 3LE, United Kingdom}

\author[0000-0002-2728-4053]{Anthony Yeates}
\affil{Department of Mathematical Sciences \\
Durham University \\
Durham DH1 3LE, United Kingdom}

\begin{abstract}
Formed through magnetic field shearing and reconnection in the solar corona, magnetic flux ropes are structures of twisted magnetic field, threaded along an axis.
Their evolution and potential eruption are of great importance for space weather.
Here we describe a new methodology for the automated detection of flux ropes in simulated magnetic fields, based on fieldline helicity.
More robust than previous methods, it also allows the full volume extent of each flux rope structure to be determined.
Our \textit{Flux Rope Detection and Organization} (FRoDO) code is publicly available, and measures the magnetic flux and helicity content of pre-erupting flux ropes over time, as well as detecting eruptions.
As a first demonstration the code is applied to the output from a time-dependent magnetofrictional model, spanning 1996 June 15 - 2014 February 10.
Over this period, 1561 erupting and 2099 non-erupting magnetic flux ropes are detected, tracked, and characterized.
For this particular model data, erupting flux ropes have a mean net helicity magnitude of $(2.66 \pm 6.82) \times 10^{43}$~Mx$^2$, while non-erupting flux ropes have a significantly lower mean of (4.04 $\pm 9.25) \times 10^{42}$~Mx$^2$, although there is overlap between the two distributions.
Similarly, the mean unsigned magnetic flux for erupting flux ropes is $(4.04 \pm 6.17) \times 10^{21}$~Mx, significantly higher than the mean value of $(7.05 \pm 16.8) \times 10^{20}$~Mx for non-erupting ropes.
These values for erupting flux ropes are within the broad range expected from observational and theoretical estimates, although the eruption rate in this particular model is lower than that of observed coronal mass ejections.
In future the FRoDO code will prove a valuable tool for assessing the performance of different non-potential coronal simulations and comparing them with observations.
\end{abstract}


\section{Introduction} \label{sec:introduction}

Flux ropes are frequently defined as bundles of solar magnetic fieldlines, twisting around a common axis.
They may emerge ready-formed from the solar interior \citep{2009SSRv..144..197L}, or may form in the atmosphere through a combination of photospheric surface flows and magnetic reconnection above polarity inversion lines \citep{1989ApJ...343..971V}.
In this case, they act to store magnetic stresses as they build in the corona.
Observationally they are associated with coronal cavities above the limb \citep{2013ApJ...770L..28B} and with filament channels on the solar disk \citep{2010SSRv..151..333M}.
Erupting filaments are often seen to be twisted, and it is believed that beyond a critical quantity of twist flux rope eruptions can push magnetic field and plasma outward into the heliosphere as a coronal mass ejection \citep[CME, ][]{2006SSRv..123..251F, 2011LRSP....8....1C}.
Understanding the formation and eruption of flux ropes is therefore critical in studying and predicting space weather phenomena.

Here we present an automated methodology to identify flux ropes within three-dimensional magnetic field data cubes.
In this paper, the methodology is applied to magnetofrictional simulations of the coronal magnetic field, driven by observational magnetogram data \citep{2014SoPh..289..631Y}.
With this methodology, flux rope volumes and photospheric footprints are precisely defined so as to enable consistent solar-cycle length statistical descriptions of eruption rates, spatial distribution, magnetic flux, and magnetic helicity.
Through the several thousand model flux ropes detected over the span of this simulation, we have an excellent database to further probe the statistics of eruption.
The long-term goal of this work is to improve our ability to predict the geo-effectiveness of Earth-directed CMEs by better understanding both their origin at the Sun and their internal magnetic structure.

The task of identifying flux ropes in a three-dimensional magnetic field dataset has received relatively little attention in the literature.
One relevant study is by \cite{2016ApJ...818..148L}, who discuss efforts to model active region NOAA 11817 through an eruptive period from 2013 August 10-12.
Using nonlinear force-free field models, they are able to model a magnetic flux rope running across the polarity inversion line.
Through tracking fieldline twist values, they are able to identify and track this magnetic flux rope as a core bundle of fieldlines.
In addition, computations indicate the presence of a high value of the squashing factor $Q$ around the boundary of this flux rope structure, suggesting a potential way to identify this distinct topological region.

A rather different study was undertaken by \cite{2010SoPh..263..121Y}, who investigated the appearance of magnetic flux ropes at \textit{a priori} unknown locations within a global quasi-static model.
Following the methodology described by \cite{2009ApJ...699.1024Y}, flux ropes were detected within the volume by searching for locations with inward magnetic tension and outward magnetic pressure forces, supplemented with a criterion of minimum parallel current.
From six distinct time periods in solar cycle 23, flux ropes were detected and classified.
Major findings included the doubling of the number of simulated flux ropes from cycle minimum to maximum, with the rate of flux rope ejection increasing by a factor of eight.
The analysis was subsequently extended to a continuous simulation running throughout the period 1996--2012 \citep{2014SoPh..289..631Y}.

In this work, we extend and improve upon several aspects of these existing methodologies.
One limitation of the \cite{2009ApJ...699.1024Y} approach is that it detects only the flux rope core, or axis.
Our new methodology allows us to define the full extent of each flux rope, enabling us to more accurately measure the magnetic flux and helicity content of each flux rope over its lifetime.
In addition, the new methodology is less prescriptive of the precise geometrical shape of the magnetic field within the flux rope, providing a more robust definition alongside a computationally efficient method.
We also feel it to be more practical than the squashing-factor approach where it can be difficult to identify which of the many topological regions in a complex magnetic field correspond to flux ropes.

This paper focuses on describing the methodology itself, and on illustrating the results for the same global simulation as \cite{2014SoPh..289..631Y}.
The simulation is briefly described in \S~\ref{sec:model}, before the methodologies for detecting flux ropes and their eruption are described in \S~\ref{sec:methodology}.
The simulation is slightly extended compared to that presented in \cite{2014SoPh..289..631Y}, spanning the years 1996 through 2014.
Magnetic field data are output at a cadence of twenty-four hours, and used to generate a flux-rope database sufficient for meaningful statistics, which are summarized in \S~\ref{sec:flux-rope-properties}.
In future it is hoped that this method can be extended to compare different coronal models, in order to better understand the origin of CMEs.
To this end, our flux-rope detection code is freely available to the community.

\section{Coronal magnetic field model} \label{sec:model}

With the goal in mind of simulating solar filament channels, \cite{2000ApJ...539..983V} developed a mean field model to simulate large-scale regions of the sun.
Using this model framework, \cite{2006ApJ...641..577M} worked to develop a simulation of a portion of the coronal field, allowing two magnetic bipoles to evolve and interact.
A consequence of this interaction, in the presence of photospheric footpoint motions, is the formation and eruption of several flux ropes through the course of the simulation.
\cite{2008SoPh..247..103Y} subsequently extended this work to develop a global non-potential model, driven by photospheric observations of bipolar magnetic regions and capable of continuously evolving the coronal magnetic field over months and years.
Further advances to this code, including the addition of hyperdiffusion and a variable grid, are outlined in \cite{2014SoPh..289..631Y}.

The particular simulation used for this study is an extended run of that described by \cite{2014SoPh..289..631Y}, in which the coronal magnetic field evolves quasi-statically through magnetofriction, being driven by a surface flux transport model on the lower boundary.
The surface field evolves forward through diffusion and prescribed large-scale flows, along with the emergence of new bipolar magnetic regions.

The resulting coronal field evolves through a continuous sequence of near force-free equilibria, allowing the build up of large-scale electric currents and free magnetic energy over time.
These currents tend to become concentrated in magnetic flux ropes which form over photospheric polarity inversion lines due to flux cancellation \citep{1989ApJ...343..971V}.
When flux ropes become too strong in the model, they lose equilibrium and are ejected through the outer boundary.

The quasi-static model is not capable of following the full dynamics of these ejections, although the topological evolution of the magnetic field during eruption is found to be similar to that in full magnetohydrodynamic simulations \citep{2015JApA...36..123P}.
In addition, the analyses of \cite{2010ApJ...709.1238Y} and \cite{2014SoPh..289..631Y} suggest the number of flux rope ejections in the quasi-static model to be be significantly lower than the number of observed CMEs in any given time interval.
This arises because the global model does not follow the detailed dynamics within active regions, so that, for example, multiple eruptions from within the same active region can not be reproduced in the simulation.
When higher resolution magnetograms are used as input to drive the model, it has been shown to reproduce well the formation and eruption of flux ropes within individual active regions \citep{2014ApJ...782...71G, 2017SoPh..292...90R}.
But at present it is not possible to include such fine detail within global-scale simulations, not least because magnetogram data are not available simultaneously over the full solar surface.

For the particular model run here, the grid spacing at the equator was set at 1.875 degrees and the source surface was set to 2.5~$R_\odot$.
On the photospheric boundary, supergranular diffusion was set to 450~km$^2$s$^{-1}$, with a peak meridional flow of 11~ms$^{-1}$.
In addition, a radial outflow velocity was defined as 100~ms$^{-1}$ near the outer boundary to simulate the effects of the solar wind and to keep the magnetic field radial at that height.
Further details of the grid setup and other model parameters are given in \cite{2014SoPh..289..631Y}.

\begin{figure}[ht!]
\begin{center}
\includegraphics{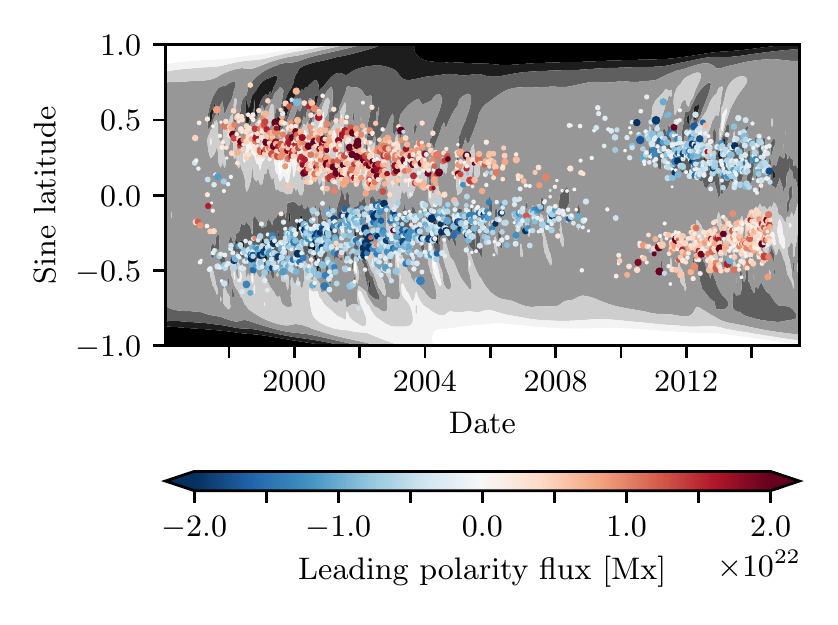}
\caption{Observed bipoles driving the magnetofrictional model, with each bipolar pair represented as a single circle colored by the magnetic flux of the leading magnetic pole.
The radius of each marker is scaled to represent the angular half-separation between magnetic peaks, with the largest marker indicating 8.69 heliographic degrees.
The mean radial magnetic field strength at 1~$R_\odot$ is plotted below this data, at levels of $\pm$\{1,3,5\}~G, in white and black, respectively.}
\end{center}
\label{fig:ars-cnf}
\end{figure}

A set of observed magnetic bipole data drives the magnetofrictional model as the source term.
The emergence time, latitude, and Carrington longitude are noted as well as the separation between magnetic peaks, magnetic flux for each polarity, tilt angle, and twist parameter.
For the work detailed here, a database of observed bipoles covers the span 15 June 1996 - 10 February 2014 \citep{10.7910/DVN/Y5CXM8}.
Figure~\ref{fig:ars-cnf} shows the observed bipoles as a function of emergence latitude and time, magnetic flux, and angular half separation.
Each bipolar region is represented as a single circle, with color mapping indicating the leading polarity magnetic flux.
Each marker is scaled to represent the angular half-separation between magnetic peaks, and background shaded contours describe the surface radial magnetic field strength.
When inserted into the three-dimensional magnetofrictional model, the bipoles take the idealized form detailed in \cite{2008SoPh..247..103Y}, with twist values distributed as outlined in \cite{2014SoPh..289..631Y}.

For analysis of the flux ropes, three-dimensional arrays of the coronal magnetic field $\vect{B}$ were stored at twenty-four hour intervals throughout the simulation, interpolated to a rectilinear grid in latitude, longitude and radius.

\section{Methodology for flux rope detection} \label{sec:methodology}

This section describes the methodology for detecting magnetic flux ropes in three-dimensional magnetic field datacubes, either individual snapshots or time series.
We have developed a set of Python routines called \textit{Flux Rope Detection and Organization} (FRoDO) that implement our methodology within a global spherical shell such as the solar corona.
This set of routines is hosted as an open source tool, in an online GitHub code repository at \url{https://github.com/lowderchris/FRoDO} \citep{FRoDO_2017_842785}.
Here we outline how the flux ropes and their eruptions are detected, illustrated using selected times from the global coronal model in \S~\ref{sec:model}.
Statistics for the full model run are then presented in \S~\ref{sec:flux-rope-properties}.

\subsection{Magnetic helicity mapping} \label{sec:magnetic-helicity-mapping}

The basic premise behind this methodology is to identify flux ropes as concentrations of high fieldline helicity in the corona.
Fieldline helicity is defined on each magnetic fieldline $L$ within the simulation domain by the line integral
\begin{equation} \label{eqn:helicity}
\mathcal{A}(L) = \int_{L(x)} \frac{\vect{A} \cdot \vect{B}}{|\vect{B}|} \, \mathrm{d}l,
\end{equation}
where $l$ represents arc length along the magnetic fieldline and $\vect{A}$ is a vector potential for the magnetic field $\vect{B}$, meaning that $\vect{B}=\nabla\times\vect{A}$.
The quantity $\mathcal{A}(L)$ has been introduced in the coronal context by \cite{2016A&A...594A..98Y}, who discuss its physical interpretation in more detail.
Essentially, $\mathcal{A}(L)$ measures the magnetic helicity in an infinitesimally thin tubular domain around the fieldline $L$.
If the footpoints of the fieldline are fixed, then $\mathcal{A}$ is an ideal invariant, just like the total magnetic helicity.
If footpoint motions are significant -- as in the large-scale corona -- then these can change the amount of fieldline helicity on coronal fieldlines.
Typically there is an overall gradual injection of helicity, which becomes concentrated in flux ropes due to reconnection \citep[e.g.][]{2016A&A...594A..98Y}.
The fieldline helicity provides a way to quantify this process precisely, and to identify the locations where most helicity is stored.

Since $\mathcal{A}(L)$ depends on the choice of $\vect{A}$, it is necessary to specify a particular gauge.
This is equivalent to specification of the reference field in the commonly-used relative magnetic helicity \citep[see][]{2014ApJ...787..100P}.
Following \cite{2016A&A...594A..98Y}, we employ the DeVore gauge where
\begin{eqnarray} \label{eqn:gauge}
\vect{A}(r,\theta,\phi) = \frac{R_\odot}{r}\vect{A}_0(\theta,\phi) + ...\nonumber \\
... + \frac{1}{r}\int_{R_\odot}^r\vect{B}(r',\theta,\phi)\times\vect{e}_rr'\,\mathrm{d}r'
\end{eqnarray}
and $\vect{A}_0(\theta,\phi)$ satisfies $\nabla\cdot\vect{A}_0=0$.
This gauge is commonly used due to its computational convenience \citep[e.g.][]{2016SSRv..201..147V}.
It is useful for our application since the second term of (\ref{eqn:gauge}) makes $\mathcal{A}(L)$ sensitive to twisting of the horizontal magnetic field with increasing $r$, which is a good indicator of flux ropes.
At the same time the first term makes $\mathcal{A}(L)$ sensitive to how well-aligned the fieldline is with contours of $B_r$ when projected on the lower boundary $r=R_\odot$.
This term also generates a significant contribution for coronal flux ropes since they tend to be aligned along photospheric polarity inversion lines rather than perpendicular to them.

\begin{figure}[ht!]
\begin{center}
\includegraphics{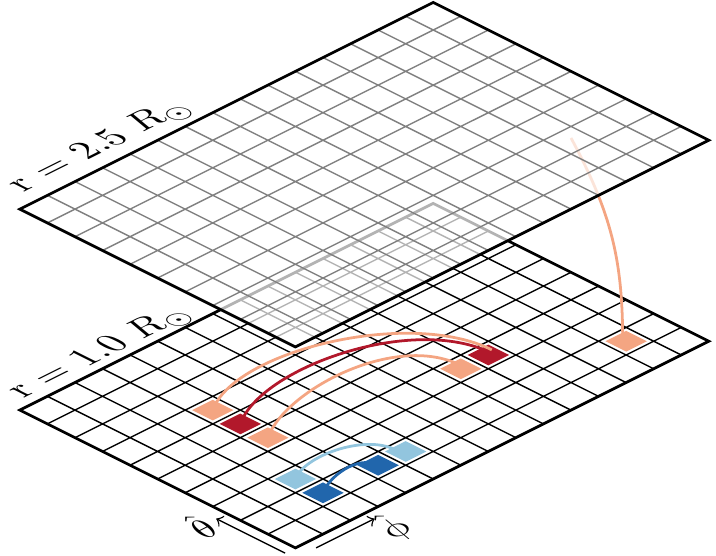}
\caption{Cartoon schematic of magnetic fieldline helicity mapping.
Open and closed magnetic fieldlines with positive (red) or negative (blue) fieldline helicity have these values recorded at both ends in an array on $r = R_\odot$.
If more than one value is recorded in a given pixel, that of largest magnitude is kept.}
\end{center}
\label{fig:plt-helicity-schematic}
\end{figure}

To begin the search for flux ropes, the fieldline helicity is mapped on an equally-spaced grid in longitude and sine-latitude on the photospheric boundary $r=R_\odot$.
To do this, we first trace magnetic fieldlines from each point in this grid and compute their fieldline helicity by integrating the fieldline helicity from Equation~(\ref{eqn:helicity}) with the vector potential from Equation (\ref{eqn:gauge}).
If the fieldline is closed (meaning that its start and end footpoints are both on the photospheric boundary) then we also assign this value to the pixel containing the end footpoint.
If a pixel is thereby assigned two or more different values of fieldline helicity, we keep the value that is largest in magnitude.
This procedure is illustrated in Figure~\ref{fig:plt-helicity-schematic}.

\begin{figure}[ht!]
\begin{center}
\includegraphics{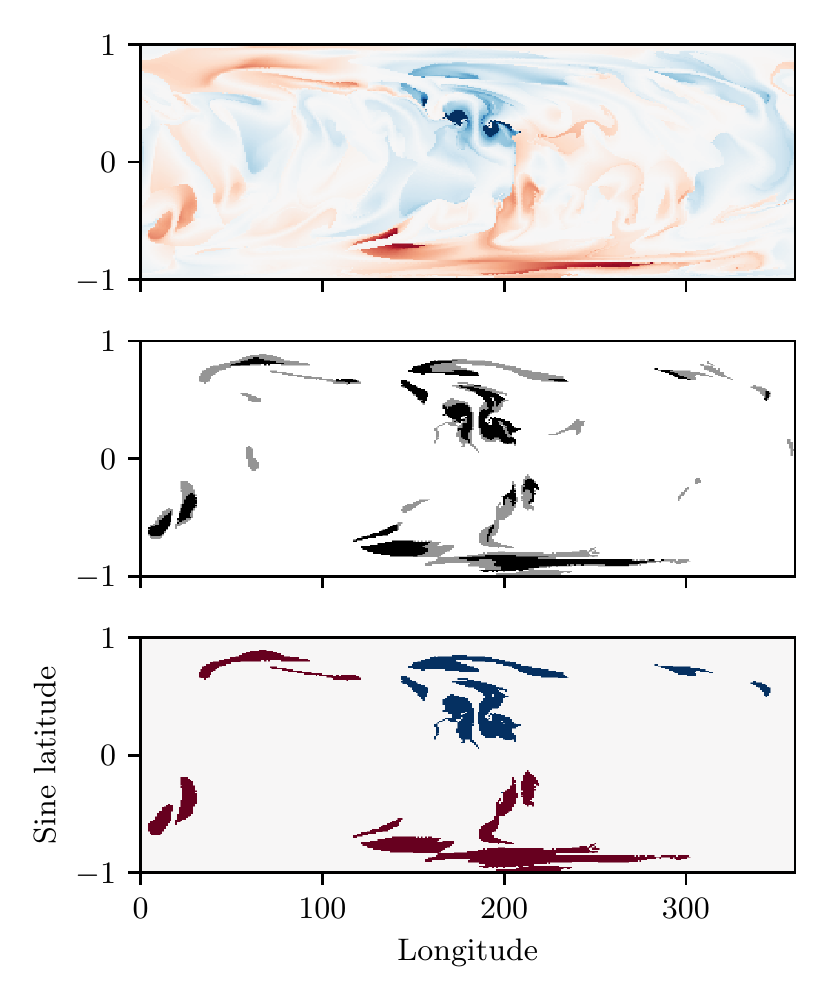}
\caption{(upper) Photospheric mapping of fieldline helicity $\mathcal{A}(\theta,\phi)$, scaled to $\pm 1.44 \times 10^{22}$~Mx, with positive helicity in red and negative in blue.
(middle) Core threshold regions are marked in black, with the full extent mapped in grey.
(lower) Final flux rope footprint map, marking the dominant helicity sign within each region.
All of the above subpanels display helicity mappings and derived quantities from the time 1998 December 5 12:00:00 UT in the coronal simulation.}
\end{center}
\label{fig:hlcy-thresholding}
\end{figure}

The result of this process is a map $\mathcal{A}(\theta,\phi)$ for fieldlines traced from a grid of fieldline footpoints on the photospheric boundary.
The upper panel of Figure~\ref{fig:hlcy-thresholding} presents one example of such a map, at the resolution of 360 pixels in longitude and 180 pixels in sine latitude.
With equally spaced pixels in both dimensions, the resulting map contains pixels of uniform physical surface area.
This same map resolution and scale is used for all of the remaining calculations.
The distribution of $\mathcal{A}(\theta,\phi)$ at the surface is marked by smooth progression at smaller scales, with the appearance of distinct domains clustered around the roots of more complex and twisted field.
Polarity inversion lines in the photospheric $B_r$ show up as lines where $\mathcal{A}(\theta,\phi)=0$, since the length of fieldlines goes to zero as the footpoints approach these lines.
In principle, similar maps could be produced at different heights in the corona; the flux rope detection described in \S~\ref{sec:flux-rope-detection} is based on the map at $r=R_\odot$, although additional maps at $r=2.5\,R_\odot$ are used for detecting eruptions (\S~\ref{sec:erupting-flux-rope-detection}).

\begin{figure}[ht!]
\begin{center}
\includegraphics{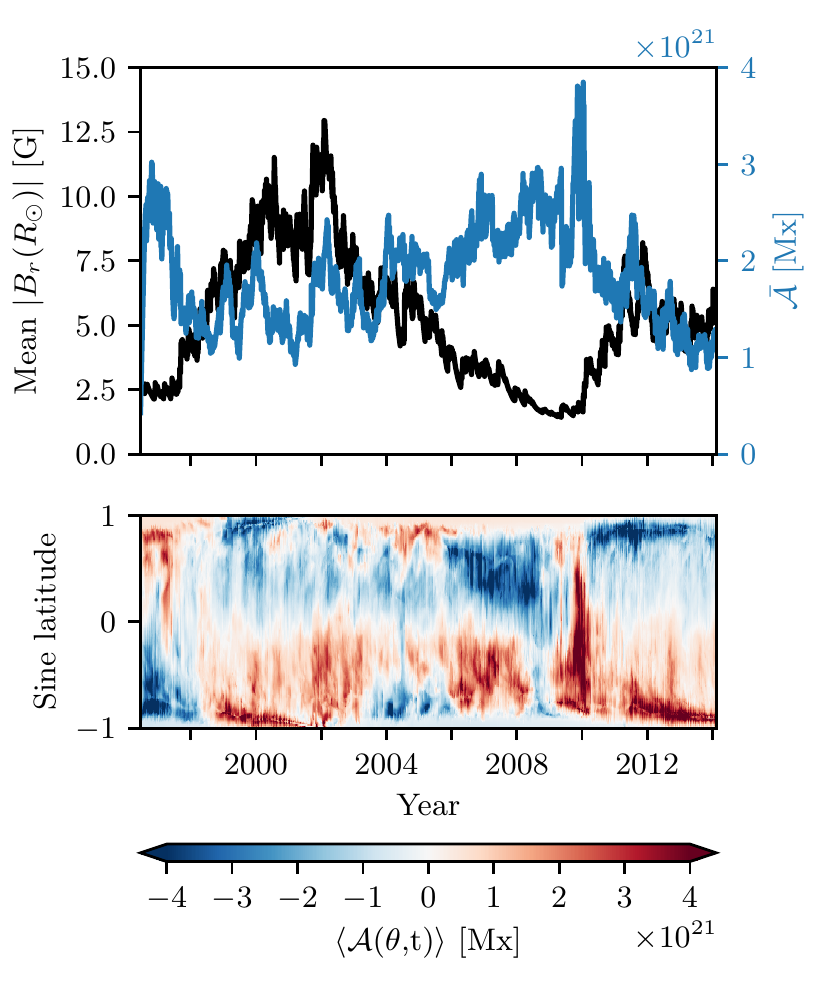}
\caption{(upper) Comparison of mean unsigned values of photospheric magnetic field strength (black) and fieldline helicity (blue).
(lower) Time-latitude profile of longitude-averaged fieldline helicity on the photosphere.}
\end{center}
\label{fig:br0_hlcy_lprofs}
\end{figure}

When mapped over the full time domain of this simulation data (1996 June 15 - 2014 February 10), global changes in the pattern of fieldline helicity are evident.
Figure~\ref{fig:br0_hlcy_lprofs} (upper) displays the mean unsigned fieldline helicity
\begin{equation}
\overline{\mathcal{A}} (t) = \frac{1}{4 \pi} \int \big| \mathcal{A}(\theta, \phi, t) \big| ~ \mathrm{d}(\cos \theta) ~ \mathrm{d} \phi
\label{eqn:habs}
\end{equation}
as a function of time in blue, along with the mean unsigned photospheric magnetic field strength in black.
It is clear that there is no direct correlation between the two quantities in this model, indicating that twisted magnetic field structures in the corona are not simply located within active regions with fieldline helicity proportional to the field strength.
Rather the topology of the coronal magnetic field is more complex and develops through gradual footpoint shearing and reconfiguring over time.

Figure~\ref{fig:br0_hlcy_lprofs} (lower) marks out the distribution of longitude-averaged fieldline helicity
\begin{equation}
\langle \mathcal{A}(\theta, t) \rangle = \frac{1}{2 \pi} \int_0^{2 \pi} \mathcal{A}(\theta, \phi, t) ~ \mathrm{d} \phi
\end{equation}
at 1~$R_\odot$ as a function of latitude and time.
With the exception of a few periods during this simulation, a hemispheric pattern emerges with negative and positive helicity dominating the northern and southern hemispheres, respectively.
This is in accordance with known hemispheric patterns of helicity on the Sun \citep{2003AdSpR..32.1867P}, and with the results of \cite{2012ApJ...753L..34Y} for this model, in which paper the pattern is seen in other diagnostics, namely chirality and current helicity density.
The origin of this pattern in the model is explained by \cite{2009SoPh..254...77Y}.
The distinct period in late 2009, where positive helicity extends far into the northern hemisphere, is addressed within \S~\ref{sec:distributions}.

\subsection{Flux rope detection} \label{sec:flux-rope-detection}

Using a map of fieldline helicity on the photospheric boundary, as in the upper panel of Figure~\ref{fig:hlcy-thresholding}, a thresholding process is applied to identify the footprints of magnetic flux ropes.
In particular, twisted flux ropes are expected to exhibit a higher magnitude of fieldline helicity than the neighboring field \citep{2016A&A...594A..98Y}.
Fieldline helicity is therefore used as the criteria to define flux rope structures in this work.

Two thresholding values are employed to map the extent of flux rope footprint boundaries, and are illustrated in Figure~\ref{fig:hlcy-thresholding}-(middle) for the sample map.
A core threshold value, $\tau_c$, defines the strong cores of flux rope structures as exhibited in their mark on the photospheric $\mathcal{A}(\theta,\phi)$ distribution.
An extent threshold value, $\tau_e$, defines the outer boundary enveloping these strong cores.
In the figure, regions with fieldline helicity magnitude greater than the core threshold, $|\mathcal{A}(\theta,\phi,t)| > \tau_c(t)$, are mapped in black.
Those with fieldline helicity magnitude exceeding the extent threshold but not the core threshold, $\tau_c(t) > |\mathcal{A}(\theta,\phi,t)| > \tau_e(t)$, are marked in grey.
The rationale of two thresholds is to enable the full extent of each flux rope structure to be identified, while at the same time excluding regions without a sufficiently twisted core.

To capture the evolution of flux rope structures over extended periods of time, we found that an adaptive set of thresholds is required since the fieldline helicity depends on magnetic field strength.
However, it is not simply a matter of scaling the thresholds according to the global unsigned magnetic flux through the photosphere, as the flux ropes are coronal structures.
We have seen in Figure~\ref{fig:br0_hlcy_lprofs} that the mean unsigned fieldline helicity is not correlated with the photospheric unsigned flux for this simulation.
Instead, we found more consistent detection of flux ropes over time if the thresholds were scaled with the mean unsigned fieldline helicity $\overline{\mathcal{A}}(t)$ defined in Equation (\ref{eqn:habs}).
Accordingly, the threshold values are scaled with the relations,
\begin{equation} \label{eqn:threshs}
\tau_c(t) = \frac{\overline{\mathcal{A}}(t)}{\overline{\mathcal{A}}_{\rm ref}} \tau_{c,{\rm ref}} ~ ; ~ \tau_e(t) = \frac{\overline{\mathcal{A}}(t)}{\overline{\mathcal{A}}_{\rm ref}} \tau_{e, {\rm ref}}.
\end{equation}
For this particular simulation, we determined suitable parameters to be $\tau_{c,{\rm ref}} = 4.84\times 10^{21}$~Mx, $\tau_{e,{\rm ref}} = 3.39\times 10^{21}$~Mx, and $\overline{\mathcal{A}}_{\rm ref} = 1.29 \times 10^{21}$~Mx.
These values for $\tau_{c,{\rm ref}}$ and $\tau_{e,{\rm ref}}$ were chosen through careful consideration of detected structures throughout various phases in the solar cycle.
The parameter $\tau_{c, {\rm ref}}$ is chosen relative to the mean fieldline helicity reference value, $\overline{\mathcal{A}}_{\rm ref}$, to provide consistent detection of flux rope core fieldlines of large helicity magnitude.
Through careful calibration, the secondary extent thresholding parameter $\tau_{e, {\rm ref}}$ was selected to provide a compact but well-defined set of fieldlines surrounding this strong core.
Reduction of the value of this extent threshold results in expanded sheaths of fieldlines surrounding flux rope cores.
Likewise an increase of this threshold reduces this set of fieldlines, paring down towards the large fieldline helicity magnitude core.
These parameters may need to be adjusted if the method is applied to data from other simulations.

Figure~\ref{fig:hlcy-thresholding}-(bottom) displays the final output of flux rope footprint regions, colored by the sign of fieldline helicity (positive in red and negative in blue).
To arrive at this final map, distinct regions are taken from the extent threshold, discarding regions without a strong core.
Detected regions with a surface area extent less than $9.38 \times 10^{18}$cm$^2$ (10 pixels at this particular resolution) are removed from consideration.
In this manner nearby strong flux rope core footprints are bridged, and isolated weaker footprints are removed.
As a final step, regions are separated by sign of fieldline helicity, such that two adjacent regions of opposite helicity sign are not merged.
The result is a labeled map of flux rope footprints, with connectivity mapping preserved for that point in time.

\begin{figure*}[ht!]
\begin{center}
\includegraphics[width=2.25in]{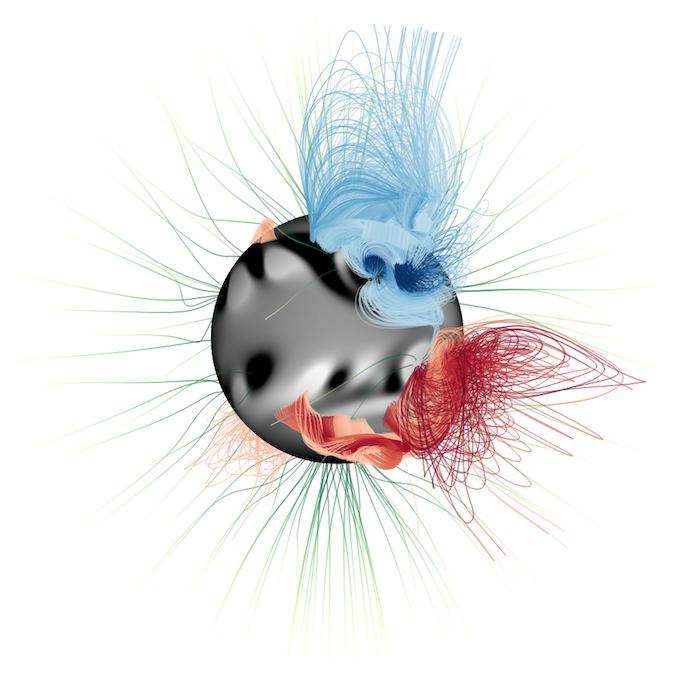}
\includegraphics[width=2.25in]{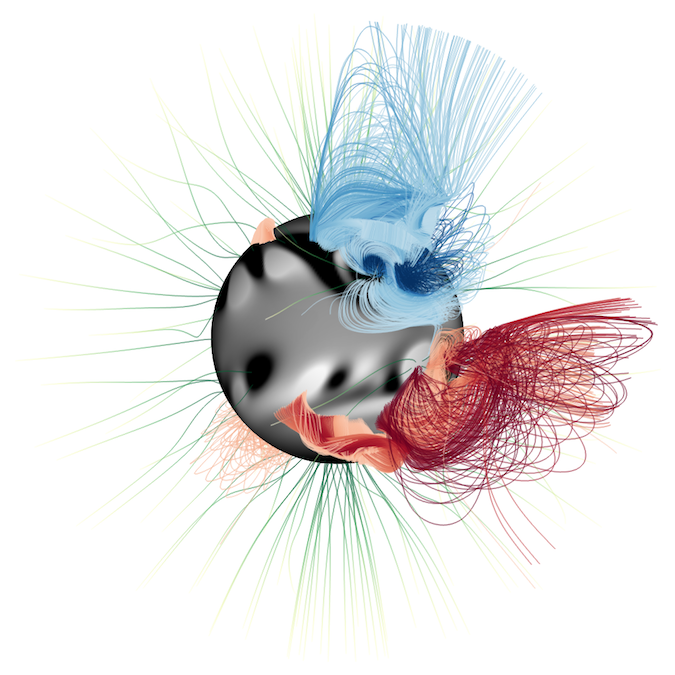}
\includegraphics[width=2.25in]{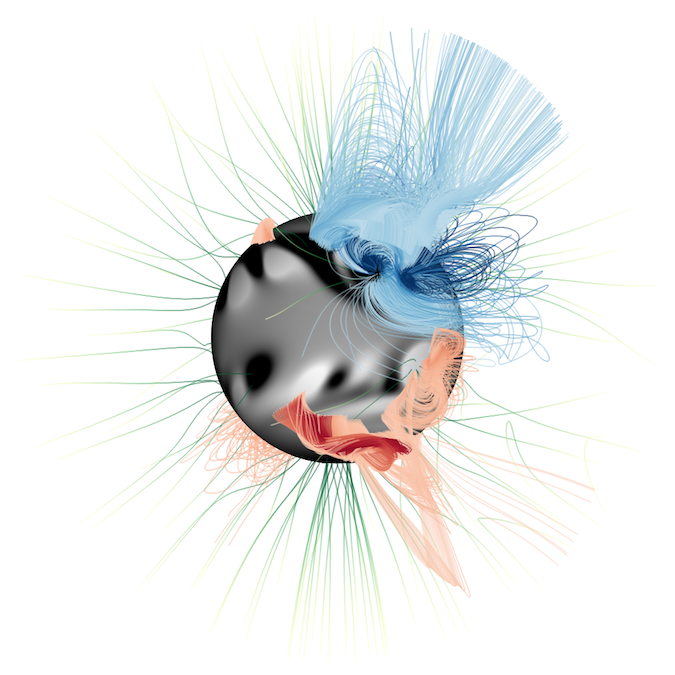}
\caption{Magnetic fieldlines of the flux ropes identified in Figure~\ref{fig:hlcy-thresholding}, colored by fieldline helicity between $\pm 1.44 \times 10^{22}$~Mx with red positive and blue negative.
The surface radial magnetic field at $r = R_\odot$ is shown between $\pm$10~G with white positive and black negative.
A selection of open magnetic fieldlines are plotted in a green-yellow color table, scaled with radius.
Times displayed are: (left) 1998 December 2 12:00:00 UT, (middle) 1998 December 3 12:00:00 UT, and (right) 1998 December 4 12:00:00 UT.
An extended animation of this sequence is available in the online version of the journal.}
\end{center}
\label{fig:hlcy-flmap-3d}
\end{figure*}

Figure~\ref{fig:hlcy-flmap-3d} shows coronal magnetic field lines traced from the flux rope footprints identified in Figure~\ref{fig:hlcy-thresholding}, colored red/blue according to their fieldline helicity.
For context, the yellow/green fieldlines show a selection of background open magnetic field lines outside the identified flux ropes.

It is important to note that the identified flux rope field lines may have one footprint region that is much less compact than the other.
In fact, they need not even have two footprint regions identified in the fieldline helicity map.
This is because flux ropes do not exist in isolation; many of the fieldlines in these structures may be further connected to other regions of the corona and may not return to the photosphere as a single coherent bundle.
In other words, the entire length of the fieldline need not be part of the flux rope.
This is evident for some of the examples in Figure~\ref{fig:hlcy-flmap-3d}, where the full lengths of the fieldlines are plotted.
The fact that $\mathcal{A}$ is computed by integrating along the whole fieldline could mean that different coronal structures are folded into the same location in the photospheric $\mathcal{A}(\theta,\phi)$ map, but this is rare in practice, at least for the simulation considered here.

Note that the present analysis utilizes the magnetic fieldline helicity as a quantity for the detection and tracking of flux ropes.
Prior attempts at tracking these features utilized integrated fieldline parallel current.
Appendix~\ref{sec:parallel-current-thresholding} outlines some of these earlier efforts, and problems therewith.

\subsection{Flux rope tracking} \label{sec:flux-rope-tracking}

The flux rope detection process is repeated for each timeframe in the simulation under consideration, providing snapshots of flux rope footpoint locations, and their associated magnetic fieldlines.
These snapshots are linked in time, searching through prior and subsequent frames for sufficient footpoint overlap (greater than 50\% overlap in area) to identify flux ropes from one frame to the next.
The result is a dataset of flux rope footprints and fieldlines, and their time histories, uniquely labeled over the course of the simulation.

With these established time histories, two final criteria are placed on these structures.
First, tall arcades of magnetic fieldlines may have sufficient values of $\mathcal{A}$ due to fieldline length to register as initial features.
To remove these the maximum radial extent is computed for each fieldline within each potential structure.
These values are then averaged to provide a mean maximum radial extent for each structure, and the evolution of that value for the history of that structure.
Features that spend more than half of their lifetime with a mean maximum radial extent above 1.25~$R_\odot$ are removed.
Secondly, detected flux rope structures with only a single day of duration are considered spurious, and are removed.
With these features removed, the resulting database contains only flux rope structures.

\subsection{Erupting flux rope detection} \label{sec:erupting-flux-rope-detection}

\begin{figure}[ht!]
\begin{center}
\includegraphics[width=3.0in]{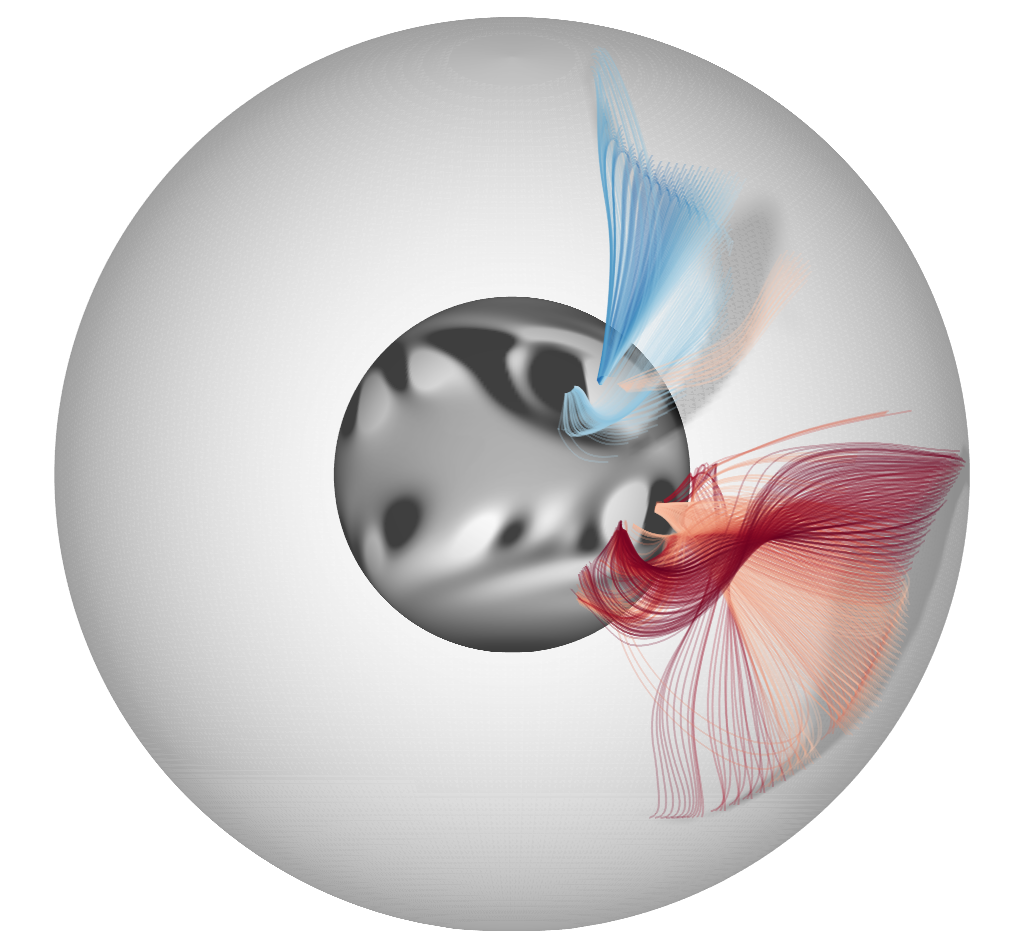}
\caption{Example of erupting flux rope structures.
The surface radial magnetic field strength is shown at $R_\odot$, scaled from $\pm$10~G in white and black.
A translucent surface at 2.5~$R_\odot$ displays the horizontal magnetic field strength at that location, with larger magnitudes in shades of increasing darkness.
Detected erupting magnetic fieldlines are traced down from this surface and colored according to fieldline helicity, ranging from $\pm 1.44 \times 10^{22}$~Mx, indicated in red and blue.
The time displayed is 1998 December 3 12:00:00 UT.
An animated version of this eruption diagram is available in the online version of the journal.}
\end{center}
\label{fig:erupt-diagram-snp}
\end{figure}

From the perspective of space weather, particular importance is attached to those flux ropes that erupt.
To identify which flux ropes in the database eventually erupt through the outer boundary of the domain, it is not enough simply to look for their sudden disappearance.
In the simulation, flux ropes can lose equilibrium and erupt through the outer boundary, but they can also disappear through reconnection if the magnetic field is reconfigured, often due to new flux emergence.
Moreover they occasionally fall below the detection threshold either through coronal diffusion or because of the time-varying nature of the threshold itself.
To ensure that no false eruptions are detected, our approach is to independently identify eruptions through the outer boundary at $2.5\,R_\odot$, and then to trace these down to the photosphere to associate with pre-existing flux rope footprints in the database.
Since the simulation data are processed at a finite cadence (once per day), it is possible that eruptions are missed by this approach.
But the timescales of eruption in the global magnetofrictional model are longer than in the real corona due to the quasi-static nature of the evolution, so this cadence is sufficient to capture the majority of events.

To search for flux rope eruptions through the outer boundary at $2.5\,R_\odot$ for a particular time, maps are made of both the fieldline helicity and of the horizontal magnetic field $B_\perp := (B_\theta^2 + B_\phi^2)^{1/2}$ on this boundary \citep[cf.][]{2016A&A...594A..98Y}.
As a flux rope structure migrates through the upper simulation boundary as a part of the eruption process, these two quantities should provide a unique signature of these structures.
The orientation of erupting flux ropes leads to a strong signal in the magnitude of horizontal magnetic field strength at this boundary, providing an excellent identification flag.
Adding an additional threshold on field line helicity doubly ensures that this signature is linked with a detected flux rope, and provides a clean method for linking with pre-erupting signatures.
For each step in time under consideration, the mean value of the horizontal magnetic field strength at a radius of 2.5~$R_\odot$ is computed as $\bar{B}_{\perp}$.
Using a reference comparison value of $\bar{B}_{\perp,\textrm{ref}}$ = 0.0276~G, a thresholding value is defined as ($\bar{B}_{\perp} / \bar{B}_{\perp,\textrm{ref}}) \cdot 0.10$~G.
This threshold value therefore scales with variations in the horizontal magnetic field strength over the course of the solar cycle.
Candidate regions are then identified where $B_\perp$ is greater than this thresholding value for that particular time.
Then, the mapping of fieldline helicity is searched to ensure that this detected candidate region contains fieldlines above the original thresholding parameters $\tau_c(t)$ and $\tau_e(t)$.
For regions with significant horizontal magnetic field strength as well as overlap with fieldlines of sufficient fieldline helicity, a positive identification of an erupting flux rope is made.
From these detected signature points at 2.5~$R_\odot$, magnetic fieldlines are traced down to their photospheric end-points.
These end-points are then compared with the footprints of detected flux ropes from the previous time step.
If this eruption signature is linked to a flux rope footprint in this manner, that flux rope is labeled as eruptive.

Figure~\ref{fig:erupt-diagram-snp} displays a snapshot of two erupting flux ropes in the simulation at 1998 December 3 12:00:00 UT.
The outer translucent surface at 2.5~$R_\odot$ displays the horizontal magnetic field strength at this radius, with larger magnitudes of this value in increasingly darker shades of grey.
From detection at this outer boundary the resulting associated flux rope fieldlines are traced out, with color shading indicating fieldline helicity values ranging $\pm 1.44 \times 10^{22}$~Mx in red and blue, respectively.
Two large erupting flux rope structures are clearly visible, one in each hemisphere.
These two flux ropes are subsequently compared with the previous snapshot in time, linking their signatures with their pre-eruptive histories.

\section{Flux rope properties} \label{sec:flux-rope-properties}

With a full database of flux rope positional histories throughout the domain of this simulation, their properties can be explored.
Through eruption tracking, detected flux ropes are labeled as either erupting or non-erupting.
These two populations are separated in the following analysis.

\subsection{Spatial and temporal distributions} \label{sec:distributions}

\begin{figure}[ht!]
\begin{center}
\includegraphics{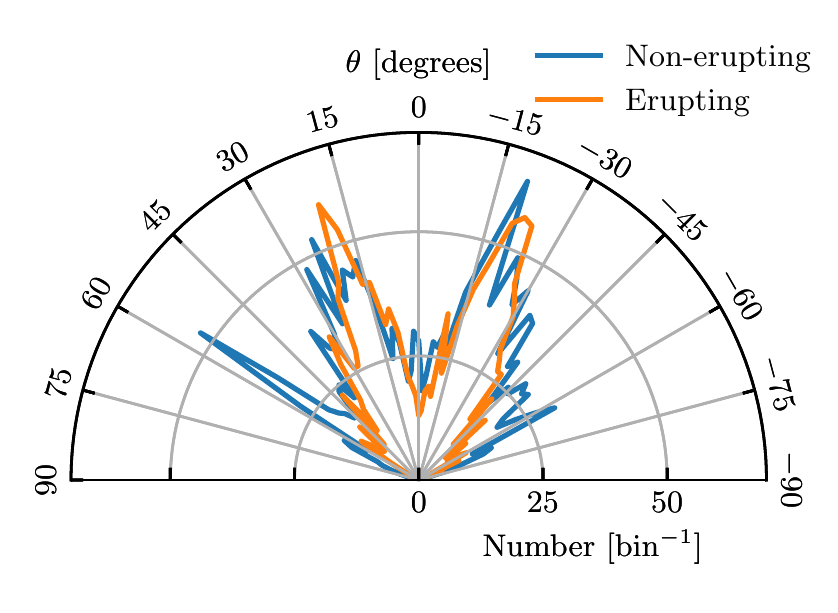}
\caption{Angular distribution of flux rope mean latitude for erupting (orange) and non-erupting (blue) flux ropes.
Flux ropes are sorted within bins of angular size 2 degrees.}
\end{center}
\label{fig:fr-mlats}
\end{figure}

Figure~\ref{fig:fr-mlats} shows the angular distribution of detected flux ropes, as a function of latitude.
For each of the flux rope time histories, the mean footprint latitude is computed at the moment of maximum footprint-enclosed unsigned magnetic flux.
Typically this is near the end of the flux rope's lifetime, since the majority of flux ropes grow gradually over time.
In total 1561 erupting flux ropes and 2099 non-erupting flux ropes were detected over the course of the simulation.
While both distributions show two primary mid-latitudinal peaks, the non-erupting ropes are more prevalent than erupting ropes at higher latitudes, up to about $\pm$60~degrees, and also at the equator.
In contrast, erupting flux ropes are more highly concentrated around the mid-latitudinal peaks, centered on about $\pm$20~degrees.
This is consistent with the results of \cite{2014SoPh..289..631Y}, who found that flux rope eruption rates were greater from active latitudes, despite a larger fraction of the volume being filled by flux ropes outside of active latitudes.

\begin{figure*}[ht!]
\begin{center}
\includegraphics{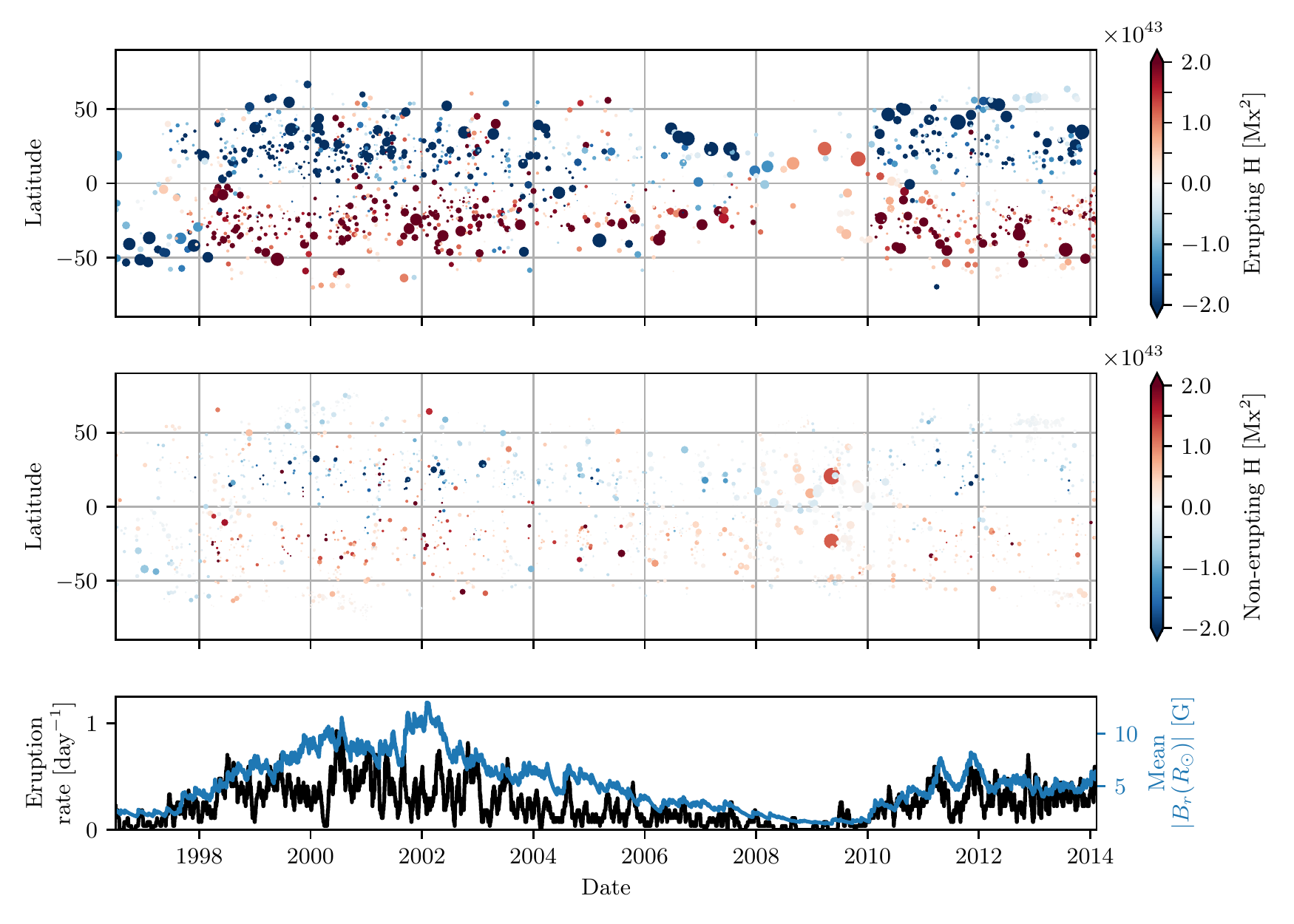}
\caption{(upper) Distribution of \textit{non-erupting} model flux rope footprint mean latitudes in time, captured at maximum enclosed unsigned magnetic flux.
The radius of each glyph indicates the relative area of each footprint, with a maximum footprint area of $4.26 \times 10^{21}$~cm$^2$.
The color mapping indicates the helicity of each flux rope, scaled from $\pm2 \times 10^{43}$~Mx$^2$ in red (positive) and blue (negative).
(middle) An identical mapping of \textit{erupting} flux ropes.
(lower) Measure of the 27-day averaged detected flux rope ejection rate, as a function of time.
The mean surface radial magnetic field strength is displayed in blue.}
\end{center}
\label{fig:fr-bfly-nhlcy-erupt}
\end{figure*}

To analyze this distribution in more detail, Figure~\ref{fig:fr-bfly-nhlcy-erupt} shows how the erupting (upper) and non-erupting (middle) flux ropes are distributed in both latitude and time.
As in Figure~\ref{fig:fr-mlats}, the flux rope properties are calculated at the moment of maximum footprint-enclosed unsigned magnetic flux.
Each glyph's radius marks the relative footprint area $A$, with a maximum of $4.26 \times 10^{21}$~cm$^2$.

A first clear pattern here is that flux ropes are limited within certain latitudinal bounds, and these bounds move closer to the equator at solar minimum.
This reflects the fact that flux ropes form above polarity inversion lines on the photosphere, and follow the distribution of those lines as the solar cycle progresses.
This distribution is also evident in observations of solar filaments \citep{1994A&A...290..279M}.
Moreover, we see that the larger glyphs -- indicating larger footpoint areas -- tend to occur outside of active latitudes.
This is because these flux ropes have had longer to grow in size due to photospheric shearing, in comparison to the younger structures found at active latitudes.

The color mapping in Figure~\ref{fig:fr-bfly-nhlcy-erupt} indicates the helicity contained within each flux rope, displayed between $\pm2 \times 10^{43}$~Mx$^2$ in red and blue, respectively.
This is computed from the fieldline helicity map on the photospheric boundary and is given by
\begin{equation}
H = \int_{A}\mathcal{A}(\theta,\phi)|B_r(\theta,\phi)|\,R_\odot^2 ~ \mathrm{d} (\cos \theta) ~ \mathrm{d}\phi,
\end{equation}
where the integral is taken over the flux rope's footprint region $A$ \citep[cf.][]{2016A&A...594A..98Y}.
If a flux rope has two identified footprint regions in the photosphere, the one with the largest footprint area is chosen.
Two trends are most obvious here: a tendency for erupting ropes to have more helicity than non-erupting ropes (to be analyzed in \S~\ref{sec:comparison-properties}), and a tendency for negative helicity in the northern hemisphere and positive in the southern hemisphere.
This hemispheric pattern reflects the overall distribution of field line helicity seen in Figure~\ref{fig:br0_hlcy_lprofs}, allowing for the fact that individual flux ropes can be exceptions to the hemispheric pattern \citep{2009SoPh..254...77Y}.
The pattern holds both for erupting and non-erupting flux ropes, and independently of their size and strength.
Indeed the pattern is observed \textit{in situ} in magnetic clouds at 1~AU \citep{1994GeoRL..21..241R}.

Finally, the (lower) panel of Figure~\ref{fig:fr-bfly-nhlcy-erupt} shows, in black, a running 27-day average of the flux rope eruption rate.
To compare with the overall evolution of magnetic field, the mean surface radial magnetic field strength is plotted in blue.
The flux rope ejection rate follows broadly alongside the evolution of this field, waxing and waning with the solar cycle.
This behavior is consistent with \cite{2014SoPh..289..631Y}.
However, the detected eruption rate, with an overall mean of 0.24 per day, is rather lower than shown in Figure 6 of \cite{2014SoPh..289..631Y}, where the mean is 0.49 per day.
This discrepancy is found to be the result of the previous methodology erroneously detecting features that the current methodology would not classify as flux ropes.
In other words, the new criterion is more stringent.

\begin{figure}[ht!]
\begin{center}
\includegraphics[width=3.00in]{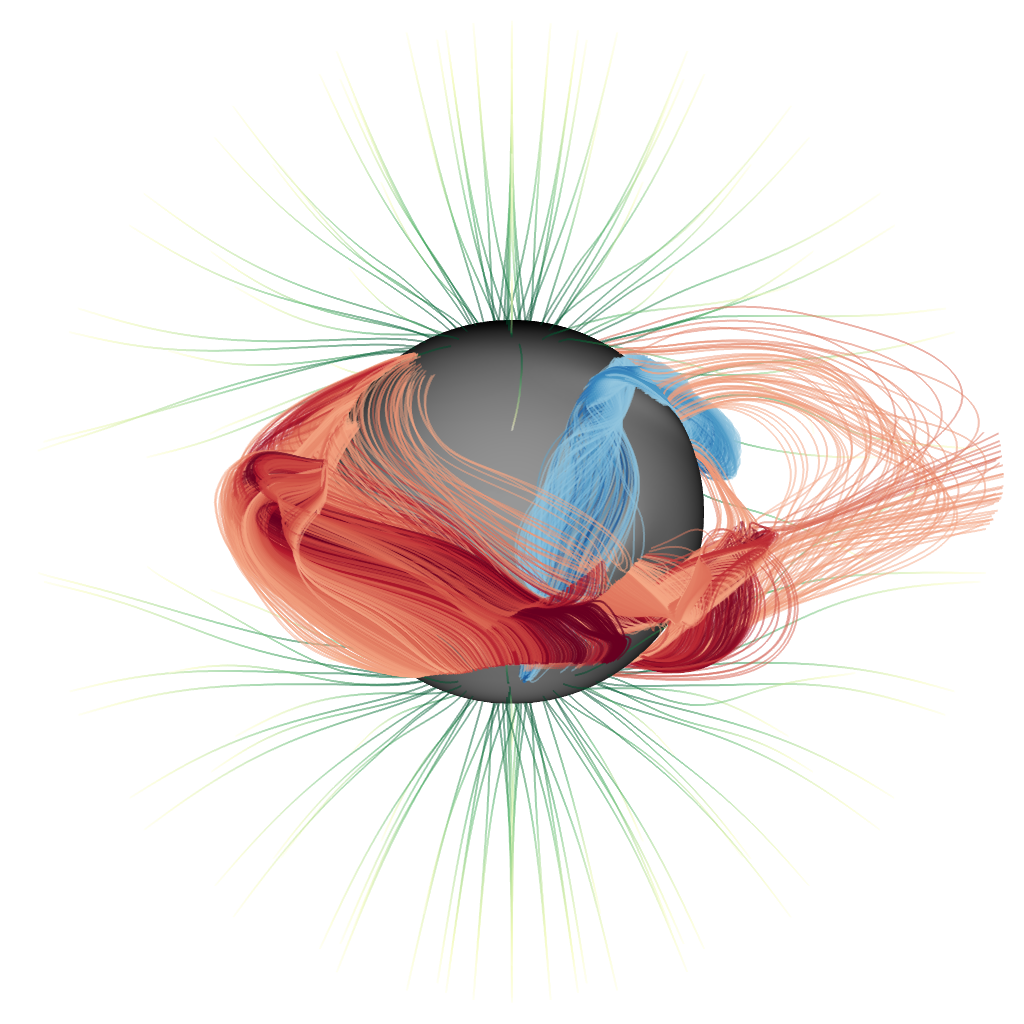}
\caption{An example snapshot during the extended Minimum period, specifically for 2009 August 29 12:00:00 UT.
The format is the same as in Figure \ref{fig:hlcy-flmap-3d}.}
\end{center}
\label{fig:hlcy-flmap-3d-solmin}
\end{figure}

The extended Minimum period from 2008-2010 is interesting because it breaks the general pattern evident in Figure~\ref{fig:fr-bfly-nhlcy-erupt}.
In particular, there are more non-erupting flux ropes with larger footprint area.
There are actually very few eruptions during this period, a trend also seen in the observed LASCO CME catalogue \citep{2009EM&P..104..295G}.
The abundance of large structures is understandable given the lack of new flux emergence; the existing polarity inversion lines are longer-lived, and the magnetic field structure is also on a larger-scale.
The large spatial extent of these flux ropes allows them to build up high net fieldline helicity, but it is relatively distributed through their volume, and is not sufficiently concentrated to cause them to lose equilibrium.
The relatively weak magnetic field strengths throughout this period dampen the resulting net helicity values for these structures.
Figure \ref{fig:hlcy-flmap-3d-solmin} shows a typical example of one of these large structures, that remains stable at a low height in the corona for many days.
Because several of these large structures connect across the equator, the hemispheric helicity pattern is disrupted during this time.
The particular regions involved lead to a predominance of positive helicity, but this is probably not a systematic rule.

\subsection{Comparison of properties} \label{sec:comparison-properties}

\begin{figure}[ht!]
\begin{center}
\includegraphics{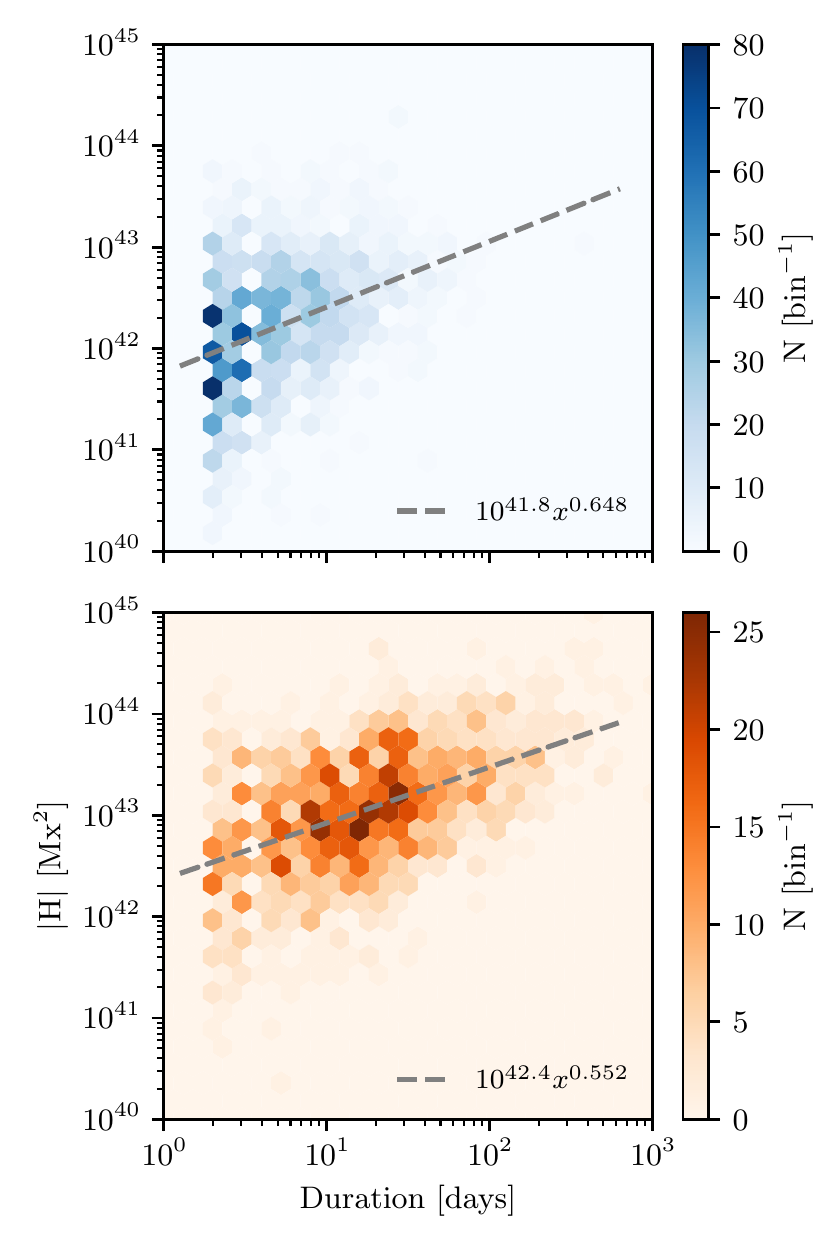}
\caption{Two dimensional histograms of the distribution of net helicity magnitude and lifetime duration for non-eruptive (blue) and eruptive (orange) flux ropes.
Of particular note is the stark shift to higher durations for eruptive flux ropes.}
\end{center}
\label{fig:fr-sct-dur-nhlcy}
\end{figure}

\begin{figure}[ht!]
\begin{center}
\includegraphics{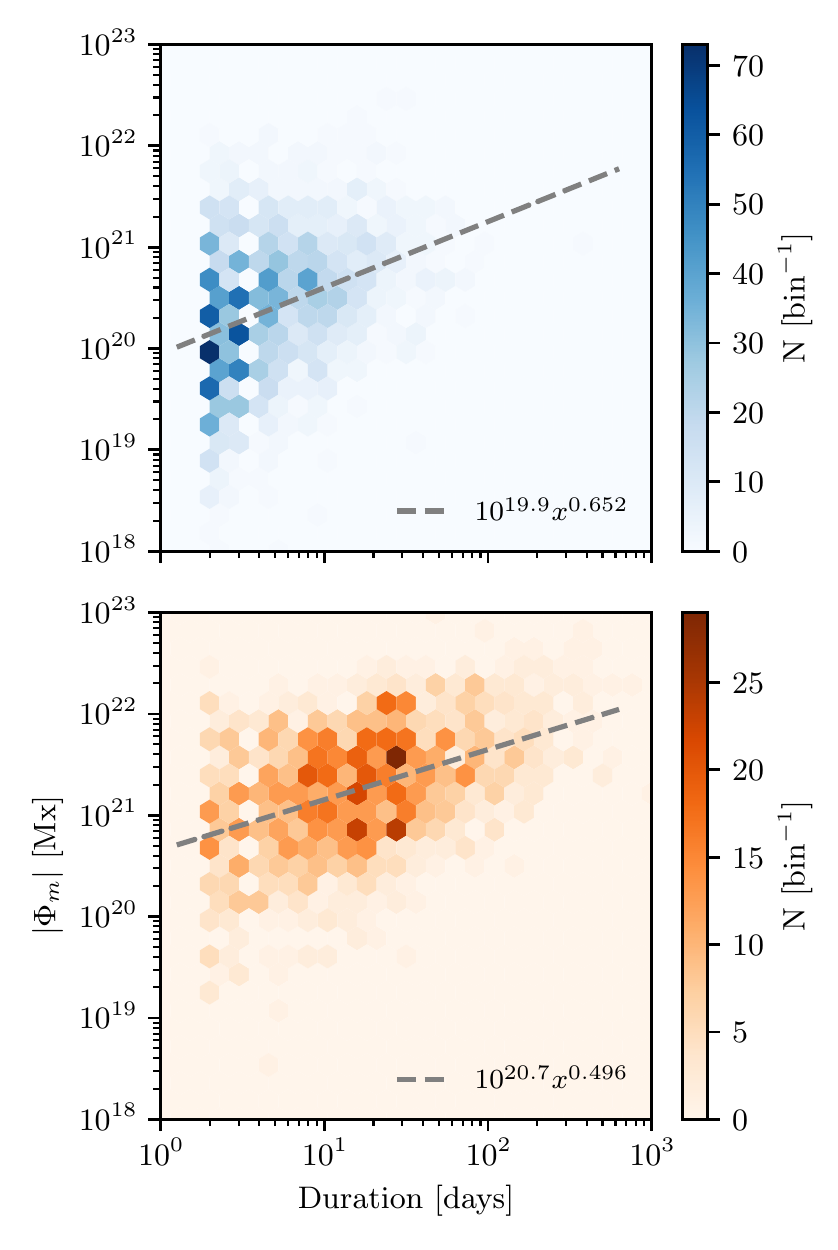}
\caption{Two-dimensional histograms of the distribution of unsigned magnetic flux and lifetime duration for non-eruptive (blue) and eruptive (orange) flux ropes.
Note the shift to higher values of enclosed unsigned magnetic flux for eruptive flux ropes.}
\end{center}
\label{fig:fr-sct-dur-uflux}
\end{figure}

\begin{figure}[ht!]
\begin{center}
\includegraphics{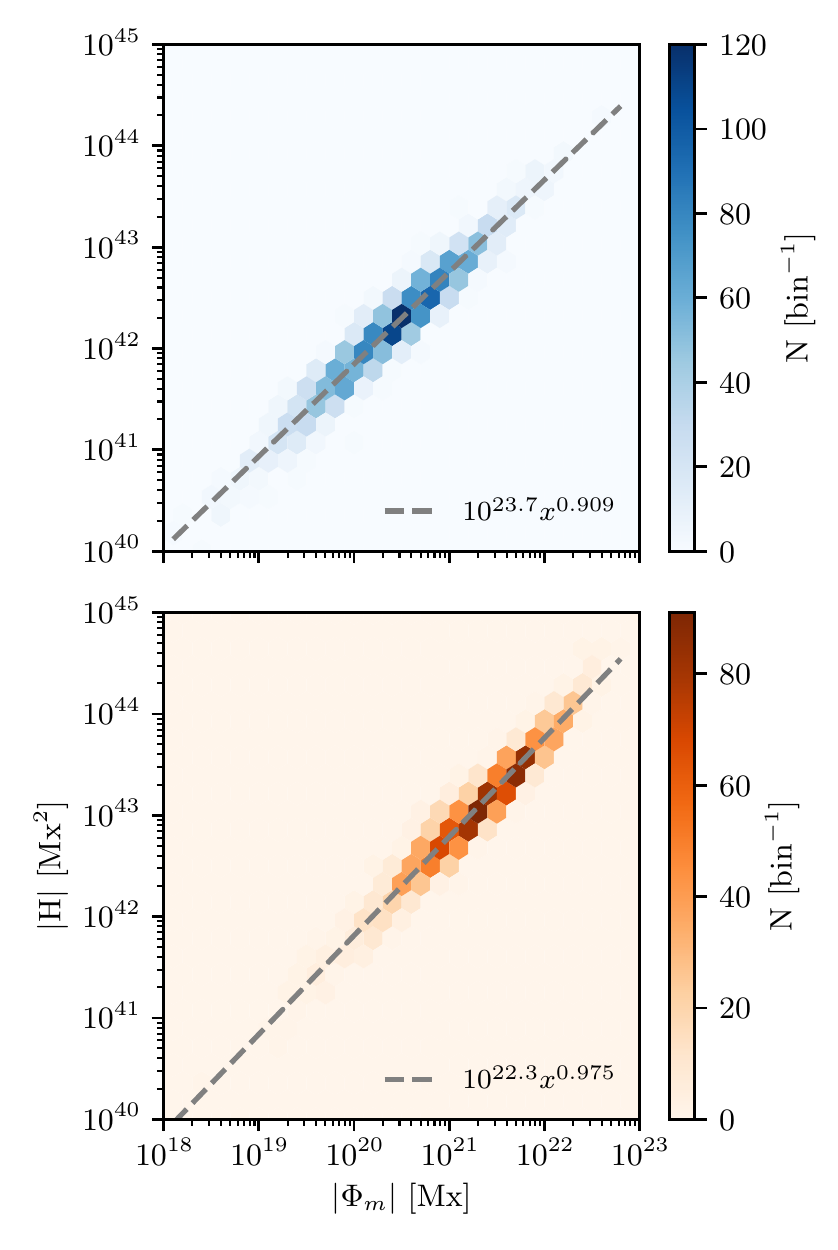}
\caption{Two-dimension histograms of the distribution of net helicity magnitude and unsigned magnetic flux for non-eruptive (blue) and eruptive (orange) flux ropes.
Note the relative shift to larger values of unsigned magnetic flux and net helicity magnitude for eruptive flux ropes.}
\end{center}
\label{fig:fr-sct-uflux-nhlcy}
\end{figure}

Figures \ref{fig:fr-sct-dur-nhlcy} to \ref{fig:fr-sct-uflux-nhlcy} show two-dimensional histograms of helicity versus lifetime, flux versus lifetime, and helicity versus flux, respectively, for the full set of flux ropes.
In each case, the non-erupting and erupting populations are separated in order to assess any differences between them.
As in Figure~\ref{fig:fr-bfly-nhlcy-erupt}, magnetic flux and helicity for each flux rope are computed at the time of maximum footprint-enclosed magnetic flux, and each erupting rope is assigned values of these quantities from its largest area footprint (if it has two).
The resulting two-dimensional distributions are binned into hexagonal bins in a log-log space.
For each distribution, a power law is fitted, plotted in dashed grey with the determined functional form indicated.
Note for the distributions of flux rope duration a minimum duration of two days has been imposed, as detailed earlier.

\begin{deluxetable}{cccll}
\tablecaption{Flux rope statistics}
\tablecolumns{5}
\tablenum{1}
\label{tab:fluxrope_stats}
\tablewidth{0pt}
\tablehead{
\colhead{} & \multicolumn{2}{c}{Quantity} & \multicolumn{2}{c}{Spearman}\\
\cline{2-3} \cline{4-5}\\
\colhead{E/NE} & \colhead{1} & \colhead{2} & \colhead{cc} & \colhead{$p$-value}}
\startdata
\decimals
NE & t & $|$H$|$ & 0.37 & 2.6$\times 10^{-70}$ \\
E & t & $|$H$|$ & 0.47 & 4.1$\times 10^{-85}$ \\
NE & t & $|\Phi_m|$ & 0.35 & 1.7$\times 10^{-62}$ \\
E & t & $|\Phi_m|$ & 0.41 & 1.4$\times 10^{-64}$ \\
NE & $|\Phi_m|$ & $|$H$|$ & 0.97 & 0.0 \\
E & $|\Phi_m|$ & $|$H$|$ & 0.97 & 0.0 \\
\enddata
\end{deluxetable}

Table~\ref{tab:fluxrope_stats} shows the Spearman's rank correlation coefficient for each of the six histograms.
In all cases there is a significant (low $p$-value) positive correlation, strongest for flux against helicity but still significant for helicity and flux against duration.
This supports the picture that it is the continued concentration of helicity in magnetic flux ropes over many days that often leads to the eventual eruption of flux ropes in this model.


\begin{deluxetable*}{rllll}
\tablecaption{Mean flux rope parameters}
\tablecolumns{5}
\tablenum{2}
\label{tab:fluxrope_params}
\tablewidth{0pt}
\tablehead{
\colhead{Quantity} & \colhead{Erupting} & \colhead{Non-erupting} & \colhead{t-statistic} & \colhead{$p$-value}}
\decimals
\startdata
$|\textrm{H}|$ (Mx$^2$)	& (2.66 $\pm 6.82) \times 10^{43}$	& (4.04 $\pm 9.25) \times 10^{42}$ & 13.0 & 9.17 $\times 10^{-37}$\\
${\Phi}_m$ (Mx)	& (4.04 $\pm 6.17) \times 10^{21}$	& (7.05 $\pm 16.8) \times 10^{20}$ & 20.8 & 8.29 $\times 10^{-86}$\\
${\textrm{A}}$ (cm$^2$)	& (3.57 $\pm 4.78) \times 10^{20}$	& (1.34 $\pm 2.00) \times 10^{20}$ & 17.3 & 8.34 $\times 10^{-63}$ \\
${\tau}$ (days)	& 37.3 $\pm 76.2$	& 7.00 $\pm 11.1$ & 15.6 & 5.15 $\times 10^{-51}$ \\
Number of ropes	& 1561		& 2099 & . & . \\
\enddata
\end{deluxetable*}

Next, Table~\ref{tab:fluxrope_params} shows the mean and standard deviation of each property, including also the footprint area $A$ (also at time of maximum magnetic flux).
Once again, the erupting and non-erupting populations are separated.
For each of the four properties, the t-test for independent samples (Welch's test) was calculated.
The computed t-statistic gives a measure of the separation between the means of each distribution, divided by the square root of the sum of the ratios of standard deviation squared to sample size, for each distribution.
The two-tailed p-values for all of the quantities under consideration are well below a one-percent threshold, and as such the null hypothesis of equal averages can be rejected.
Thus, the subsets of erupting and non-erupting flux ropes have statistically different distributions of net helicity magnitude, magnetic flux, footprint area, and lifetime.
Namely, erupting ropes are on average larger, longer-lasting and have higher magnitudes of helicity and magnetic flux at their peak.

The values of these properties may also be compared with those in the literature.
The duration of flux ropes in the corona cannot be determined directly from observations.
However, our mean of 37~days is consistent with the time taken for flux ropes to form and erupt in magnetofrictional simulations of simplified magnetic configurations \citep{2006ApJ...641..577M}.
This timescale depends on the coronal diffusion in the model, which was set based on the rough observational constraint that flux ropes should form above the internal polarity inversion line of a single bipolar region with around one turn of twist every 27 days.

For the flux and helicity content of erupting ropes, \textit{in situ} observations of magnetic clouds can provide useful insight, as these are understood to be formed by the eruption of flux ropes from the Sun.
Since only limited measurements are available while a spacecraft flies through a given magnetic cloud, there are considerable uncertainties on these observational estimates.
Nevertheless, it is possible to estimate the total magnetic flux and helicity content, typically by fitting a simple linear force-free magnetic flux rope to the data.
The fitted magnetic flux is typically in the range $10^{19}-10^{22}$~Mx \citep{2005JGRA..110.8107L}, and indeed the mean flux in our erupting flux ropes is close to that of these observational estimates.
Interestingly, by hypothesizing that all of the 11-year variation in interplanetary magnetic field strength is caused by CME flux, and considering the observed CME rate, \cite{2007GeoRL..34.6104O} also arrive at an average magnetic flux of $10^{21}$~Mx for each magnetic cloud.

The fitted helicity for magnetic clouds is not very well constrained since it also depends on the assumed length of the flux rope in the heliosphere.
Nevertheless, our mean of (2.66 $\pm 6.82) \times 10^{43}$~Mx$^2$ is close to that of \cite{2005JGRA..110.8107L}, if perhaps slightly lower.
A lower estimate of $2 \times 10^{42}$~Mx$^2$ was arrived at by \cite{2000ApJ...539..944D}.
So we conclude that the flux and helicity of a typical erupting flux rope in this model are reasonable and consistent with magnetic cloud observations, notwithstanding the considerable spread in flux rope properties.

\subsection{Magnetic flux and helicity ejection rates} \label{sec:ejected}

\begin{figure}[ht!]
\begin{center}
\includegraphics{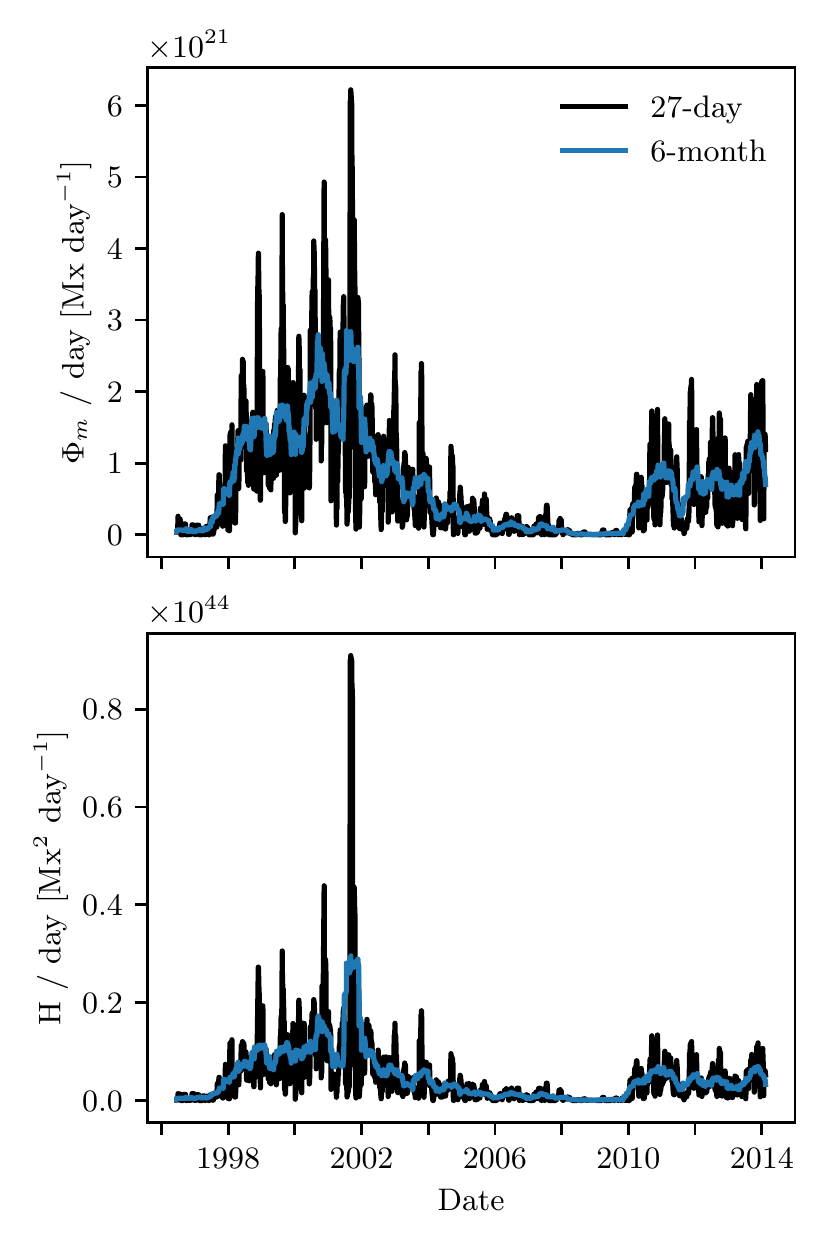}
\caption{Measures of unsigned magnetic flux (upper) and helicity (lower) ejected through the boundary at 2.5~$R_\odot$ through detected flux ropes.
The 27-day and 6-month running average values are marked in black and blue, respectively.}
\end{center}
\label{fig:fr_ejr}
\end{figure}

Having considered the properties of individual flux ropes, it is interesting to compute estimates of the overall ejection rates of magnetic flux and helicity.
Figure \ref{fig:fr_ejr} shows the ejection rates of these two quantities over the course of the simulation.
These were computed using the flux and helicity contained within each rope at its time of maximum magnetic flux, always before the ejection.
With output data at a 1-day cadence, it would not be accurate enough just to measure the flux of these quantities through the outer boundary at 2.5~$R_\odot$.
Ropes caught mid-ejection may only partially intersect this outer boundary, providing an underestimation of these ejected values.
Through this process of detecting and then identifying the history of the source flux rope, a more accurate value is provided.
It is possible that the flux and/or helicity of a rope could change by reconnection in the corona during the eruption process.
So these should be treated as estimates only.

From Figure \ref{fig:fr_ejr} it is notable that the ejected magnetic flux and helicity both vary in phase with the solar cycle.
They are also quite well correlated with one another, as would be expected from Figure \ref{fig:fr-sct-uflux-nhlcy}.

These values may be compared to previous estimates in the literature.
For example, \cite{2016SoPh..291..531D} incorporate data from 107 well-observed magnetic clouds, combined with an improved fitting model, to estimate the ejected flux and helicity.
Their estimates correspond to average ejection rates of $7.4 \times 10^{20}$~Mx~day$^{-1}$ and $6.3 \times 10^{42}$~Mx$^{2}$~day$^{-1}$.
Over the span of the simulation considered here, average ejection rates of magnetic flux and helicity in our simulation were $7.2 \times 10^{20}$~Mx~day$^{-1}$ and $4.8 \times 10^{42}$~Mx$^{2}$~day$^{-1}$.
In other words, the flux ropes in this model eject a comparable amount of magnetic flux but slightly less helicity.
However, the observed values remain rather uncertain.

Integrating over model flux rope ejection rates for Solar Cycle 23, from August 1996 to December 2008, a total of $3.5 \times 10^{24}$~Mx of magnetic flux and $2.4 \times 10^{46}$~Mx$^2$ of magnetic helicity are ejected in model flux rope eruptions.
This is close to the \citet{2016SoPh..291..531D} estimate corresponding to a total helicity ejection over Cycle 23 of approximately $2.5\times 10^{46}$~Mx$^2$.
However, \citet{1995ApJ...453..911B} estimate the lower value of $2\times 10^{45}$~Mx$^{2}$ per cycle, using data from Cycles 20 to 22, while \cite{2000ApJ...539..944D} estimates $10^{46}$~Mx$^2$ for Cycle 21.
And by considering the global-scale differential rotation, \cite{2000JGR...10510481B} estimate a total helicity ejection of $2 \times 10^{45}$~Mx$^2$ over the course of a typical solar cycle.

Note that our model, due to its large-scale nature, does not include smaller ejected structures, although \cite{2016SoPh..291..531D} find that the dominant contributions to both magnetic flux and helicity come from the larger magnetic clouds, despite them being much less numerous than smaller flux ropes.

\section{Conclusions} \label{sec:discussion-and-conclusions}

This paper has established a method for the automated detection of magnetic flux ropes in three-dimensional magnetic field simulations of the solar corona.
It has been implemented in spherical geometry in the \textit{Flux Rope Detection and Organization} (FRoDO) routine, and builds on previous work by using fieldline helicity to define the spatial extent of each flux rope.
The rope's magnetic flux and helicity may then be tracked over a time series of simulation frames.
In addition, a more robust technique for identifying flux rope eruptions has been identified and the resulting eruptions are linked back to the database of pre-erupting ropes.
The FRoDO code is freely available online \citep{FRoDO_2017_842785}.

As a first application, we have determined flux ropes in a magnetofrictional simulation of the global corona over the timespan 1996 June 15 to 2014 February 10.
This is the same simulation that was presented by \citet{2014SoPh..289..631Y}.
In total, 1561 erupting and 2099 non-erupting flux ropes were detected over the course of this simulation, for the chosen thresholds and selection criteria.
The spatiotemporal distributions of ropes and eruptions are consistent with those identified by \citet{2014SoPh..289..631Y}, but the new technique gives significantly more information about the flux rope properties.
A first broad finding is that the erupting flux ropes are, on average, larger and longer-lived than those that do not erupt, and have a larger net helicity magnitude.
These properties are consistent with the formation of these structures by gradual surface shearing and flux cancellation at photospheric polarity inversion lines \citep{1989ApJ...343..971V}.
Similarly, we recover the hemispheric helicity pattern previously seen indirectly through chirality and current helicity \citep{2012ApJ...753L..34Y}.
However, the flux ropes are rather varied both in size and morphology.
Although erupting ropes have a higher helicity and unsigned magnetic flux in general, this is not true in every individual case, and it is certainly not possible to predict whether a flux rope will erupt based on simple thresholding of these quantities.
This is in accordance with other studies, such as \citet{2017A&A...601A.125P} who considered the role of helicity, associated quantities, and ratios thereof in predicting the eruptive nature of a particular model flux rope.

Although the FRoDO code has been applied here to a single coronal simulation, it will in future provide a valuable tool for comparing the output of non-potential coronal simulations -- both against observations and against one another.
For example, the model tested here predicts a breakdown of the hemispheric helicity pattern during the weak Solar Minimum between Cycle 23 and Cycle 24, owing to the presence of large flux rope structures crossing the equator.
Verifying or disproving this behavior observationally would be an important test of this coronal model since the Minimum period is dominated by surface shearing motions with little new flux emergence.

Another finding is that sufficient magnetic flux and helicity are contained within the erupting flux ropes (before eruption) to explain the estimated ejection rate through CMEs.
This was not necessarily expected as the large-scale nature and low resolution of magnetic input data to the present model mean that it produces too few flux rope eruptions compared to observations \citep{2014SoPh..289..631Y}.
In fact, our detected eruption rate is now even lower than that using the method of \citet{2014SoPh..289..631Y}, owing to the more stringent definition of what constitutes a flux rope.
It remains to be determined in future whether the model is over-estimating the amount of flux and helicity contained in individual regions, or whether this is being released in single larger flux ropes rather than multiple smaller structures.
It is also possible that we are over-estimating the amount of flux and helicity injected, since our estimates do not account for losses caused by reconnection during the eruption process itself.
This is a topic for future study, although it may require a full MHD treatment to accurately account for the reconnection.
At present, the uncertainties in observed flux and particularly helicity content in magnetic clouds remain large enough that they do not provide a strong constraint on models.
In any case, it will likely be necessary to compare individual events rather than overall statistics, since the variation of properties (in the model at least) is large.
It is hoped that the upcoming space missions Solar Orbiter and Parker Solar Probe will provide new, stronger constraints.

In future, the FRoDO code can be applied to higher resolution models of the coronal evolution that are currently under development \citep[e.g.,][]{2016ApJ...823...55W}.
Similarly it could be applied to more hypothetical simulations of other stars \citep[e.g. that of ][]{2016MNRAS.456.3624G} or other epochs of solar activity \citep[e.g., ][]{2015ApJ...802..105R}.

\acknowledgments

This work was supported by a grant from the US Air Force Office of Scientific Research in the Basic Research Initiative ``Understanding the Interaction of CMEs with the Solar-Terrestrial Environment'', and by an STFC consortium grant to the universities of Dundee and Durham.
The authors thank C.~Prior, M.~Weinzierl, P.~Wyper, and T.~Whitbread for useful discussion concerning the development of this work, which made use of the Hamilton HPC Service of Durham University.
SOLIS data used here are produced cooperatively by NSF/NOAO and NASA/LWS.

\vspace{5mm}
\facilities{SOLIS}

\software{Python, SunPy \citep{2015CS&D....8a4009S}, SciPy \citep{2001NA....00..00J}, NumPy \citep{2001CSE...13..22W}, Mayavi \citep{2011CSE...13..40V}, Matplotlib \citep{2007CSE.....9...90H}, FRoDO \citep[][Codebase: \url{https://github.com/lowderchris/FRoDO}]{FRoDO_2017_842785}}

\appendix

\section{Parallel current thresholding} \label{sec:parallel-current-thresholding}

\begin{figure}[ht!]
\begin{center}
\includegraphics{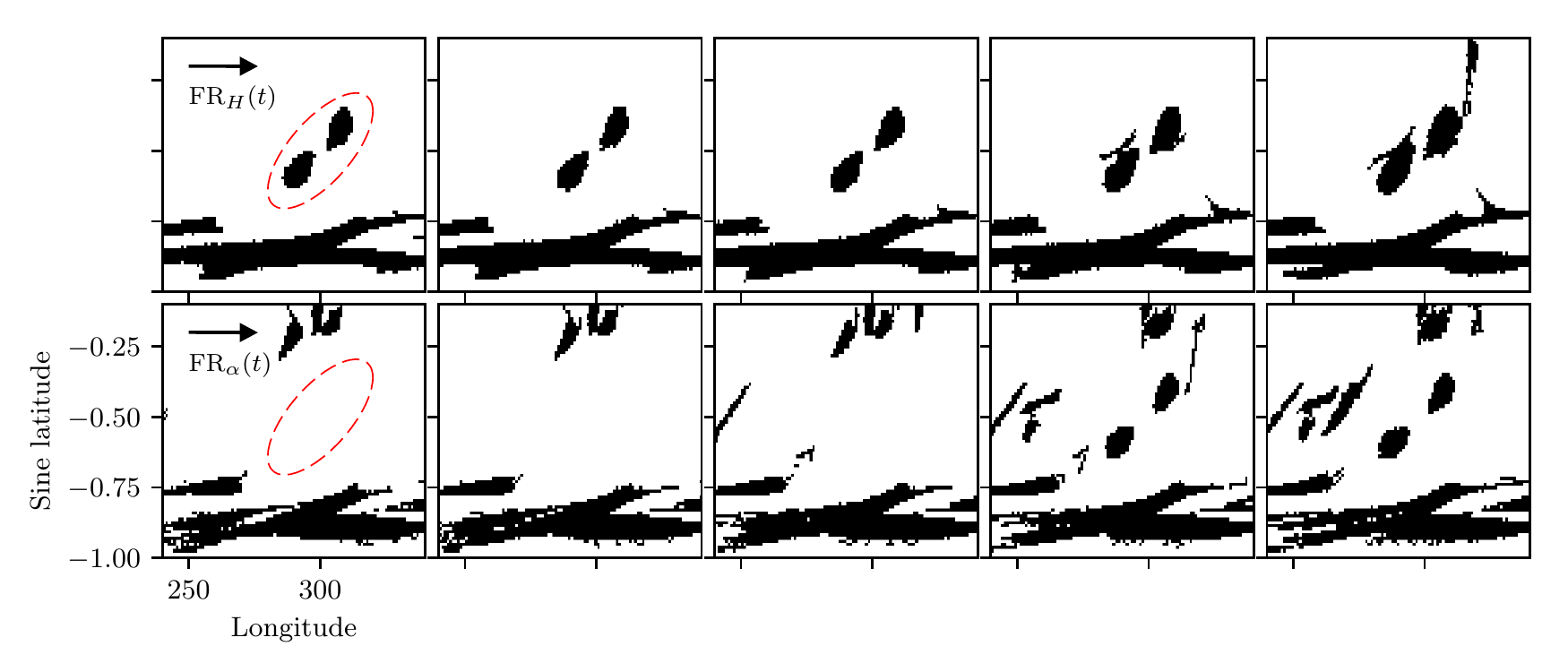}
\caption{Comparison of flux rope footpoint maps computed with fieldline helicity (upper row) and parallel current (lower row).
The sequence progresses in time from left to right, in increments of one simulation day.
One pair of flux rope footprints (marked with dashed red ellipses) is relatively well tracked in the fieldline helicity mapping, with less reliable results from parallel current tracking.}
\end{center}
\label{fig:mcomp}
\end{figure}

Prior to working with fieldline helicity as a thresholding tool for the detection of flux ropes, computed parallel current of fieldlines was considered as a potential tool.
Here, the parallel current integrated along a magnetic fieldline is defined as,
\begin{equation} \label{eqn:alpha}
\mathcal{\alpha}(L) = \int_{L(x)} \frac{\vect{J} \cdot \vect{B}}{|\vect{B}|^2} \, \mathrm{d}l,
\end{equation}
where $\vect{j}=\nabla\times\vect{B}$.
In a methodology similar to that for helicity (detailed in \S~\ref{sec:magnetic-helicity-mapping}), fieldline net parallel current was computed and mapped down to footpoints at 1.0~$R_\odot$.
From this map, thresholding techniques (as in \S~\ref{sec:flux-rope-detection}) were attempted to determine the footpoints of fieldline bundles with large magnitudes of parallel current.
However, the distribution with time proved too volatile as a stable measure for thresholding.
The result was detected flux ropes flickering in and out of detection with time, with more severe fluctuations in parallel current, which was found to be a less robust measure than fieldline helicity.
In addition, by attempting to hone thresholding values to map edges of flux rope footprints, problems propagate outward.
Many regions appear without a clear flux rope structure when visualizing magnetic fieldlines.
Fieldline helicity provided a much more stable alternative.
In comparison, the distribution of fieldline helicity appears much more clear cut, with regions more distinctly defined in time and space.
This is illustrated in Figure~\ref{fig:mcomp}.

Figure~\ref{fig:mcomp} compares a time sequence of flux rope footpoint maps computed with helicity and parallel current as a threshold.
The top row displays detected flux rope footpoints using helicity as a thresholding parameter, with the bottom row using parallel current.
These sets of maps progress in time from left to right in increments of one day.
A location of interest is marked using a dashed red ellipse.
While two distinct flux rope footpoints are visible within the helicity mapping for all five frames, the parallel current maps do not clearly resolve these same features.
These footpoints exist below the thresholding criteria for the first three frames, eventually appearing at a reduced extent.
One additional problem of parallel current thresholding is visible here.
While similar footpoint structures are detected, a host of additional regions are also classified as flux rope footpoints, which do not resemble flux ropes in fieldline tracing.

\bibliographystyle{aasjournal}
\bibliography{FRoDO}

\begin{thebibliography}{}
\expandafter\ifx\csname natexlab\endcsname\relax\def\natexlab#1{#1}\fi
\providecommand{\url}[1]{\href{#1}{#1}}

\bibitem[{{B{\c a}k-St{\c e}{\'s}licka} {et~al.}(2013){B{\c a}k-St{\c
  e}{\'s}licka}, {Gibson}, {Fan}, {Bethge}, {Forland}, \&
  {Rachmeler}}]{2013ApJ...770L..28B}
{B{\c a}k-St{\c e}{\'s}licka}, U., {Gibson}, S.~E., {Fan}, Y., {et~al.} 2013,
  \apjl, 770, L28

\bibitem[{{Berger} \& {Ruzmaikin}(2000)}]{2000JGR...10510481B}
{Berger}, M.~A., \& {Ruzmaikin}, A. 2000, \jgr, 105, 10481

\bibitem[{{Bieber} \& {Rust}(1995)}]{1995ApJ...453..911B}
{Bieber}, J.~W., \& {Rust}, D.~M. 1995, \apj, 453, 911

\bibitem[{{Chen}(2011)}]{2011LRSP....8....1C}
{Chen}, P.~F. 2011, Living Reviews in Solar Physics, 8, 1

\bibitem[{{D{\'e}moulin} {et~al.}(2016){D{\'e}moulin}, {Janvier}, \&
  {Dasso}}]{2016SoPh..291..531D}
{D{\'e}moulin}, P., {Janvier}, M., \& {Dasso}, S. 2016, \solphys, 291, 531

\bibitem[{{DeVore}(2000)}]{2000ApJ...539..944D}
{DeVore}, C.~R. 2000, \apj, 539, 944

\bibitem[{{Forbes} {et~al.}(2006){Forbes}, {Linker}, {Chen}, {Cid}, {K{\'o}ta},
  {Lee}, {Mann}, {Miki{\'c}}, {Potgieter}, {Schmidt}, {Siscoe}, {Vainio},
  {Antiochos}, \& {Riley}}]{2006SSRv..123..251F}
{Forbes}, T.~G., {Linker}, J.~A., {Chen}, J., {et~al.} 2006, \ssr, 123, 251

\bibitem[{{Gibb} {et~al.}(2014){Gibb}, {Mackay}, {Green}, \&
  {Meyer}}]{2014ApJ...782...71G}
{Gibb}, G.~P.~S., {Mackay}, D.~H., {Green}, L.~M., \& {Meyer}, K.~A. 2014,
  \apj, 782, 71

\bibitem[{{Gibb} {et~al.}(2016){Gibb}, {Mackay}, {Jardine}, \&
  {Yeates}}]{2016MNRAS.456.3624G}
{Gibb}, G.~P.~S., {Mackay}, D.~H., {Jardine}, M.~M., \& {Yeates}, A.~R. 2016,
  \mnras, 456, 3624

\bibitem[{{Gopalswamy} {et~al.}(2009){Gopalswamy}, {Yashiro}, {Michalek},
  {Stenborg}, {Vourlidas}, {Freeland}, \& {Howard}}]{2009EM&P..104..295G}
{Gopalswamy}, N., {Yashiro}, S., {Michalek}, G., {et~al.} 2009, Earth Moon and
  Planets, 104, 295

\bibitem[{{Hunter}(2007)}]{2007CSE.....9...90H}
{Hunter}, J.~D. 2007, Computing in Science and Engineering, 9, 90

\bibitem[{Jones {et~al.}(2001--)Jones, Oliphant, Peterson,
  {et~al.}}]{2001NA....00..00J}
Jones, E., Oliphant, T., Peterson, P., {et~al.} 2001--, {SciPy}: Open source
  scientific tools for {Python}, , .
\newblock \url{http://www.scipy.org/}

\bibitem[{{Lites}(2009)}]{2009SSRv..144..197L}
{Lites}, B.~W. 2009, \ssr, 144, 197

\bibitem[{{Liu} {et~al.}(2016){Liu}, {Kliem}, {Titov}, {Chen}, {Wang}, {Wang},
  {Liu}, {Xu}, \& {Wiegelmann}}]{2016ApJ...818..148L}
{Liu}, R., {Kliem}, B., {Titov}, V.~S., {et~al.} 2016, \apj, 818, 148

\bibitem[{Lowder(2017)}]{FRoDO_2017_842785}
Lowder, C. 2017, {FRoDO: Flux Rope Detection and Organization}, v0.4,  Zenodo,
  doi:10.5281/zenodo.842785.
\newblock \url{https://doi.org/10.5281/zenodo.842785}

\bibitem[{{Lynch} {et~al.}(2005){Lynch}, {Gruesbeck}, {Zurbuchen}, \&
  {Antiochos}}]{2005JGRA..110.8107L}
{Lynch}, B.~J., {Gruesbeck}, J.~R., {Zurbuchen}, T.~H., \& {Antiochos}, S.~K.
  2005, Journal of Geophysical Research (Space Physics), 110, A08107

\bibitem[{{Mackay} {et~al.}(2010){Mackay}, {Karpen}, {Ballester}, {Schmieder},
  \& {Aulanier}}]{2010SSRv..151..333M}
{Mackay}, D.~H., {Karpen}, J.~T., {Ballester}, J.~L., {Schmieder}, B., \&
  {Aulanier}, G. 2010, \ssr, 151, 333

\bibitem[{{Mackay} \& {van Ballegooijen}(2006)}]{2006ApJ...641..577M}
{Mackay}, D.~H., \& {van Ballegooijen}, A.~A. 2006, \apj, 641, 577

\bibitem[{{Mouradian} \& {Soru-Escaut}(1994)}]{1994A&A...290..279M}
{Mouradian}, Z., \& {Soru-Escaut}, I. 1994, \aap, 290, 279

\bibitem[{{Owens} {et~al.}(2007){Owens}, {Schwadron}, {Crooker}, {Hughes}, \&
  {Spence}}]{2007GeoRL..34.6104O}
{Owens}, M.~J., {Schwadron}, N.~A., {Crooker}, N.~U., {Hughes}, W.~J., \&
  {Spence}, H.~E. 2007, \grl, 34, L06104

\bibitem[{{Pagano} {et~al.}(2015){Pagano}, {Mackay}, \&
  {Poedts}}]{2015JApA...36..123P}
{Pagano}, P., {Mackay}, D.~H., \& {Poedts}, S. 2015, Journal of Astrophysics
  and Astronomy, 36, 123

\bibitem[{{Pariat} {et~al.}(2017){Pariat}, {Leake}, {Valori}, {Linton},
  {Zuccarello}, \& {Dalmasse}}]{2017A&A...601A.125P}
{Pariat}, E., {Leake}, J.~E., {Valori}, G., {et~al.} 2017, \aap, 601, A125

\bibitem[{{Pevtsov} \& {Balasubramaniam}(2003)}]{2003AdSpR..32.1867P}
{Pevtsov}, A.~A., \& {Balasubramaniam}, K.~S. 2003, Advances in Space Research,
  32, 1867

\bibitem[{{Prior} \& {Yeates}(2014)}]{2014ApJ...787..100P}
{Prior}, C., \& {Yeates}, A.~R. 2014, \apj, 787, 100

\bibitem[{{Riley} {et~al.}(2015){Riley}, {Lionello}, {Linker}, {Cliver},
  {Balogh}, {Beer}, {Charbonneau}, {Crooker}, {DeRosa}, {Lockwood}, {Owens},
  {McCracken}, {Usoskin}, \& {Koutchmy}}]{2015ApJ...802..105R}
{Riley}, P., {Lionello}, R., {Linker}, J.~A., {et~al.} 2015, \apj, 802, 105

\bibitem[{{Rodkin} {et~al.}(2017){Rodkin}, {Goryaev}, {Pagano}, {Gibb},
  {Slemzin}, {Shugay}, {Veselovsky}, \& {Mackay}}]{2017SoPh..292...90R}
{Rodkin}, D., {Goryaev}, F., {Pagano}, P., {et~al.} 2017, \solphys, 292,
  arXiv:1610.05048

\bibitem[{{Rust}(1994)}]{1994GeoRL..21..241R}
{Rust}, D.~M. 1994, \grl, 21, 241

\bibitem[{{SunPy Community} {et~al.}(2015){SunPy Community}, {Mumford},
  {Christe}, {P{\'e}rez-Su{\'a}rez}, {Ireland}, {Shih}, {Inglis}, {Liedtke},
  {Hewett}, {Mayer}, {Hughitt}, {Freij}, {Meszaros}, {Bennett}, {Malocha},
  {Evans}, {Agrawal}, {Leonard}, {Robitaille}, {Mampaey}, {Campos-Rozo}, \&
  {Kirk}}]{2015CS&D....8a4009S}
{SunPy Community}, {Mumford}, S.~J., {Christe}, S., {et~al.} 2015,
  Computational Science and Discovery, 8, 014009

\bibitem[{{Valori} {et~al.}(2016){Valori}, {Pariat}, {Anfinogentov}, {Chen},
  {Georgoulis}, {Guo}, {Liu}, {Moraitis}, {Thalmann}, \&
  {Yang}}]{2016SSRv..201..147V}
{Valori}, G., {Pariat}, E., {Anfinogentov}, S., {et~al.} 2016, \ssr, 201, 147

\bibitem[{{van Ballegooijen} \& {Martens}(1989)}]{1989ApJ...343..971V}
{van Ballegooijen}, A.~A., \& {Martens}, P.~C.~H. 1989, \apj, 343, 971

\bibitem[{{van Ballegooijen} {et~al.}(2000){van Ballegooijen}, {Priest}, \&
  {Mackay}}]{2000ApJ...539..983V}
{van Ballegooijen}, A.~A., {Priest}, E.~R., \& {Mackay}, D.~H. 2000, \apj, 539,
  983

\bibitem[{van~der Walt {et~al.}(2011)van~der Walt, Colbert, \&
  Varoquaux}]{2001CSE...13..22W}
van~der Walt, S., Colbert, S.~C., \& Varoquaux, G. 2011, Computing in Science
  \& Engineering, 13, 22

\bibitem[{Varoquaux \& Ramachandran(2011)}]{2011CSE...13..40V}
Varoquaux, G., \& Ramachandran, P. 2011, Computing in Science \& Engineering,
  13, 40

\bibitem[{{Weinzierl} {et~al.}(2016){Weinzierl}, {Yeates}, {Mackay}, {Henney},
  \& {Arge}}]{2016ApJ...823...55W}
{Weinzierl}, M., {Yeates}, A.~R., {Mackay}, D.~H., {Henney}, C.~J., \& {Arge},
  C.~N. 2016, \apj, 823, 55

\bibitem[{{Yeates}(2014)}]{2014SoPh..289..631Y}
{Yeates}, A.~R. 2014, \solphys, 289, 631

\bibitem[{Yeates(2016)}]{10.7910/DVN/Y5CXM8}
Yeates, A.~R. 2016, Bipolar magnetic regions determined from NSO synoptic
  carrington maps,  Harvard Dataverse, doi:10.7910/DVN/Y5CXM8.
\newblock \url{http://dx.doi.org/10.7910/DVN/Y5CXM8}

\bibitem[{{Yeates} {et~al.}(2010{\natexlab{a}}){Yeates}, {Attrill}, {Nandy},
  {Mackay}, {Martens}, \& {van Ballegooijen}}]{2010ApJ...709.1238Y}
{Yeates}, A.~R., {Attrill}, G.~D.~R., {Nandy}, D., {et~al.} 2010{\natexlab{a}},
  \apj, 709, 1238

\bibitem[{{Yeates} {et~al.}(2010{\natexlab{b}}){Yeates}, {Constable}, \&
  {Martens}}]{2010SoPh..263..121Y}
{Yeates}, A.~R., {Constable}, J.~A., \& {Martens}, P.~C.~H. 2010{\natexlab{b}},
  \solphys, 263, 121

\bibitem[{{Yeates} \& {Hornig}(2016)}]{2016A&A...594A..98Y}
{Yeates}, A.~R., \& {Hornig}, G. 2016, \aap, 594, A98

\bibitem[{{Yeates} \& {Mackay}(2009{\natexlab{a}})}]{2009ApJ...699.1024Y}
{Yeates}, A.~R., \& {Mackay}, D.~H. 2009{\natexlab{a}}, \apj, 699, 1024

\bibitem[{{Yeates} \& {Mackay}(2009{\natexlab{b}})}]{2009SoPh..254...77Y}
---. 2009{\natexlab{b}}, \solphys, 254, 77

\bibitem[{{Yeates} \& {Mackay}(2012)}]{2012ApJ...753L..34Y}
---. 2012, \apjl, 753, L34

\bibitem[{{Yeates} {et~al.}(2008){Yeates}, {Mackay}, \& {van
  Ballegooijen}}]{2008SoPh..247..103Y}
{Yeates}, A.~R., {Mackay}, D.~H., \& {van Ballegooijen}, A.~A. 2008, \solphys,
  247, 103

\end{thebibliography}


\end{document}